\documentclass[prb,aps,twocolumn,showpacs,floats]{revtex4-1}
\usepackage[english]{babel}
\usepackage{amsmath}
\usepackage[pdftex]{graphicx}

\begin{document}


\author{Maciej Kasperski, Henryk Puszkarski}
\affiliation{Surface Physics Division, Faculty of Physics, Adam Mickiewicz University, ul. Umultowska 85, 61-614 Poznan, Poland}
\author{Danh-Tai Hoang}
 \affiliation {%
 Asia Pacific Center for Theoretical Physics, POSTECH, San 31, Hyoja-dong, Nam-gu,
Pohang, Gyeongbuk 790-784, Korea}
\author{H. T. Diep}
\affiliation{Laboratoire de Physique Th\'eorique et Mod\'elisation, Universit\'e de Cergy-Pontoise, CNRS, UMR 8089 2, Avenue Adolphe Chauvin, F-95302 Cergy-Pontoise Cedex, France}

\title{Magnetic Properties of Two-dimensional Nanodots:  Ground State and Phase Transition}

\date{\today}
\begin{abstract}
We study the effect of perpendicular single-ion anisotropy, $-As_{\text{z}}^2$,  on the ground-state structure and finite-temperature properties of a two-dimensional magnetic nanodot in presence of a dipolar interaction of strength $D$. By a simulated annealing Monte Carlo method, we show that in the ground state a vortex core perpendicular to the nanodot plane  emerges already in the range of moderate anisotropy values above a certain threshold level. In the giant-anisotropy regime the vortex structure is superseded by a stripe domain structure with stripes of alternate domains perpendicular to the surface of the sample. We have also observed an intermediate stage between the vortex and stripe structures, with satellite regions of tilted nonzero perpendicular magnetization around the core.  At finite temperatures, at small $A$, we show by Monte Carlo simulations that there is a transition from the the in-plane vortex phase to the disordered phase characterized by a peak in the specific heat and the vanishing vortex order parameter.  At stronger $A$, we observe a discontinuous transition with a large latent heat from the in-plane vortex phase to perpendicular stripe ordering phase before a total disordering at  higher temperatures.  In the regime of  perpendicular stripe domains, namely with giant $A$, there is no phase transition at finite $T$: the stripe domains are progressively disordered with increasing $T$. Finite-size effects are shown and discussed.
\end{abstract}
\pacs{75.10.-b ; 75.40.Mg  ; 85.35.-p  ; 85.70.Kh }

\maketitle
\section{Introduction}
\label{sec:Introduction}
%
%
The ground-state (GS) structure of a magnetic nanosystem results from the competition between the interactions in the system. The energies of  the exchange and dipolar interactions as well as the anisotropy energy and the energy of interaction of the magnetic moments with the external magnetic field must be taken into account in the energy balance while determining the GS.  The combination of the frustration \cite{DiepFSS} resulting from competing interactions and the boundary effects in nanoscale systems gives rise to unexpected phenomena \cite{Diep2013}.  Among the competing forces in two dimensions (2D), let us focus on the dipolar interaction which favors in-plane spin configuration, and the perpendicular anisotropy which tends to align spins in the perpendicular axis.  The perpendicular anisotropy is known to arise with a large magnitude in ultrathin films \cite{zangwill,bland-heinrich}.  Note that, in thin films with Heisenberg and Potts models, the competing dipolar interaction and  perpendicular anisotropy causes
a spin re-orientation transition at a finite temperature \cite{Santa2000,Hoang2013}.

In this paper, we focus on the case of a ultrasmall magnetic nanodot with Heisenberg spins.   Four GS configurations have been observed in such nanosystems \cite{RochaJAP107}: (i) a capacitor-like state, (ii) a planar vortex, (iii) a vortex with core, and (iv) a domain structure. The conditions of formation of each structure have been determined \cite{RochaJAP107} with the size of the sample, the relation between the exchange and dipolar interactions, and the type of crystal order taken into account. Systems in which the core vortex structure occurs hold much promise from the commercial point of view; the occurrence of this structure has already been demonstrated experimentally \cite{Shinjo289,Raabe88,Wachowiak298} by different imaging techniques. A major advantage of core vortex structures is the central region (core) of nonzero perpendicular magnetization, the polarization of which is stable at room temperature (as shown by Shinjo \emph{et al.} \cite{Shinjo289}). Interestingly, core magnetization reversal \cite{Kikuchi90, Xiao102} can be realized in two ways, by applying a strong magnetic field \emph{perpendicular} to the surface of the sample, or a short pulse of magnetic field \emph{parallel} to it. This property of magnetic nanodots opens the door to their application in magnetoresistive random access memory (MRAM).

The current development of a technology that allows to obtain nanosamples with a very strong perpendicular anisotropy \cite{Broeder60,Kurt108,Hodumi90} has inspired us to investigate, with the use of Monte Carlo (MC) simulations \cite{Landau09,Brooks11}, the behavior of the core vortex structure, so interesting from the point of view of applications, under the impact of \emph{giant} perpendicular anisotropy.

The purpose of this work is (i) to investigate  the GS configuration in magnetic nanodots taking into account the short-range exchange interaction, the long-range dipolar interaction and the perpendicular anisotropy in 2D, (ii) to study the nature of the ordering and the phase transition at finite temperatures in such nanodots. The methods we employ in this paper are MC simulations with different techniques.

Section \ref{GSM} is devoted to the determination of the GS, while section \ref{FTB} shows MC results of finite-temperature behaviors. Concluding remarks are given in section \ref{Concl}.

\section{Ground state}\label{GSM}
\subsection{Model and method of ground-state determination}
\label{sec:Assumptions}

Let us consider a 2D system of Heisenberg spins occupying  the sites of a square
lattice within a finite $L\times L$ square. The Hamiltonian of the system is assumed to have the standard form:
\begin{eqnarray}
H &=& - J\sum_{ij}^{\text{nn}} \vec{s}_i\cdot\vec{s}_j
	- A\sum_{i}^{\text{all}}(s^{z}_i)^2\nonumber \\
   &&- D\sum_{ij}^{\text{all}}
		\left[
			\frac{3(\vec{s}_i\cdot\vec{r}_{ij})(\vec{s}_j\cdot\vec{r}_{ij})}{r_{ij}^5} - \frac{\vec{s}_i\cdot\vec{s}_j}{r_{ij}^3}
		\right],\label{eq:ham}
\end{eqnarray}
where $J$ denotes the exchange integral, $D$ is the dipolar coupling parameter, $A$ is the single-ion uniaxial perpendicular anisotropy parameter, $\vec{s}_i$ ($|\vec{s}_i|=1 \text{ for all } i$) is the spin at the $i$-th site, and $\vec{r}_{ij}$ ($r_{ij}=|\vec{r}_{ij}|$) is the position vector connecting the spins at the $i$-th and $j$-th sites. The first summation runs over all the nearest-neighbor spin pairs $ij$, the second one over all the spins in the system, and the third one over all  spin pairs.  The dipolar energy is calculated from the expression included in the Hamiltonian (\ref{eq:ham}) without any numerical approximations; in particular we do not introduce the cut-off radius, since this has been shown \cite{RochaJAP107,Vedmedenko59} to affect quantitatively the calculation results in a sensible manner.
In the following, $J$ is taken as a fixed parameter and is used as the energy unit ($J=1$). The GS and thermal properties are calculated with varying $A$ and $D$. 
\par
%
%
\par
To find the GS of the system defined above we have used the simulated annealing method of the MC simulation class \cite{Brooks11}. The main steps of the procedure are:

(i) to generate a random spin configuration, which corresponds to a high temperature phase of the system,

(ii) to update one by one all the spins as follows. At a lattice site we calculate the energy of  its spin $E_1$.
Then, we take a random spin orientation and calculate its new energy $E_2$.  If $E_2<E_1$
the new spin orientation is accepted. Otherwise, it is accepted only
with a probability $p=\exp [-(E_2-E_1)/k_BT]$ where $k_B$ is the Boltzmann constant and $T$ temperature,

(iii) to repeat the previous step a sufficient number of times,

(iv) to reduce the temperature $T$ and get back to step (ii).  With decreasing temperature, the spins converge closer and closer to the GS with iterations.

The above algorithm is thus a  ``slow cooling" procedure which works rather well for small systems without strong bond disorder. Another way to get the GS is to use the steepest-descent method: i) we generate an initial spin configuration, ii) at each lattice site we calculate the local field from other spins acting on that site, iii) we align the spin of that site along the local field to minimize its energy, iv) we take another spin and repeat step ii) until all spins are considered, iv) we iterate the procedure a large number of times until the convergence of the system energy to a minimum. In general, except for spin glasses, systems without strong bond disorder converges to the GS with a few dozens to a few hundred iterations \cite{Ngo2007}.  We have checked that this method  gives the same results as those obtained from the simulated annealing shown in the following.

\subsection{Evolution of the ground state: formation of the vortex core and stripe domains}
\label{sec:GroundStateStructures}
\begin{figure}[h]
	\centering
		\includegraphics[width=4cm]{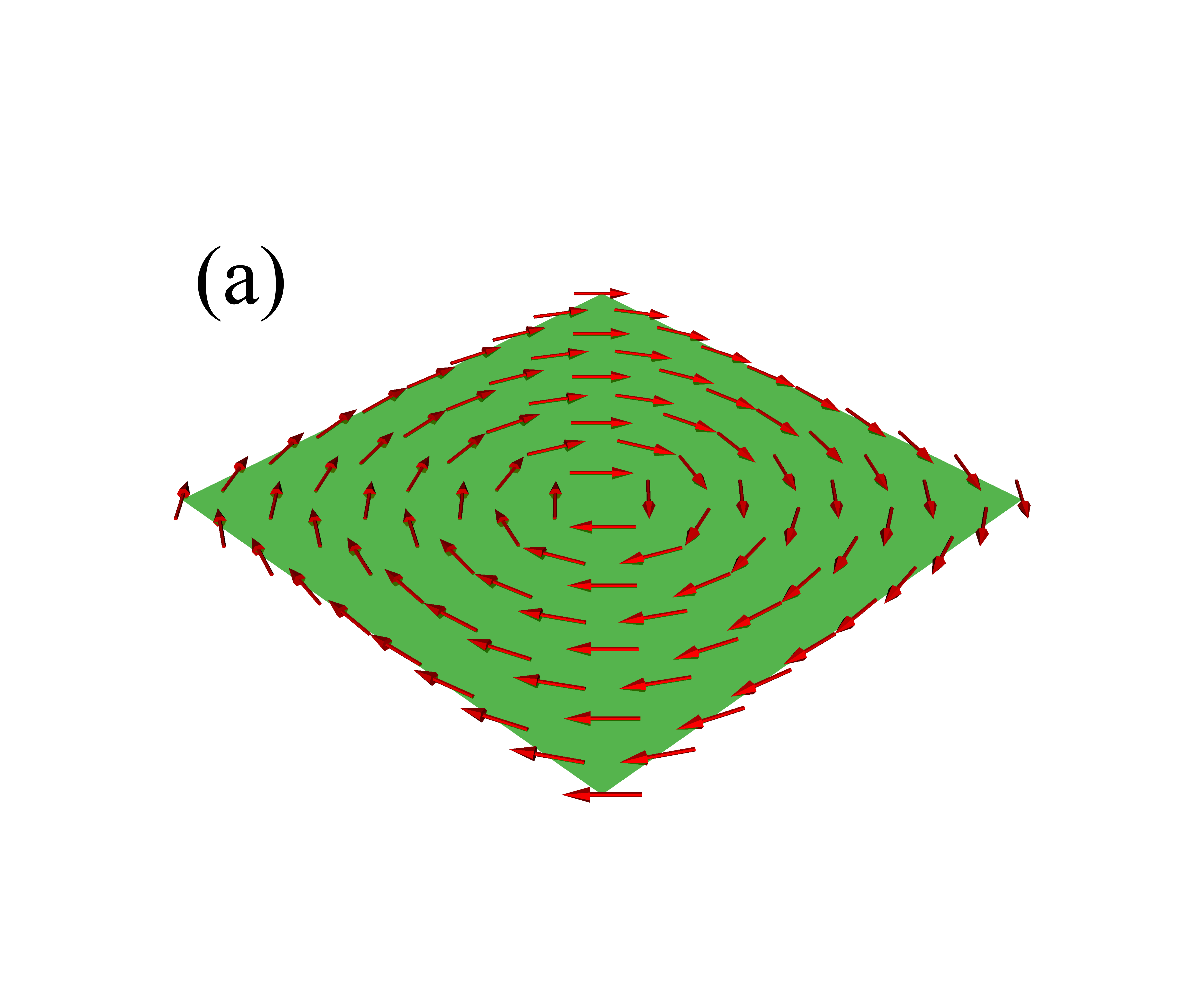}
\includegraphics[width=4cm]{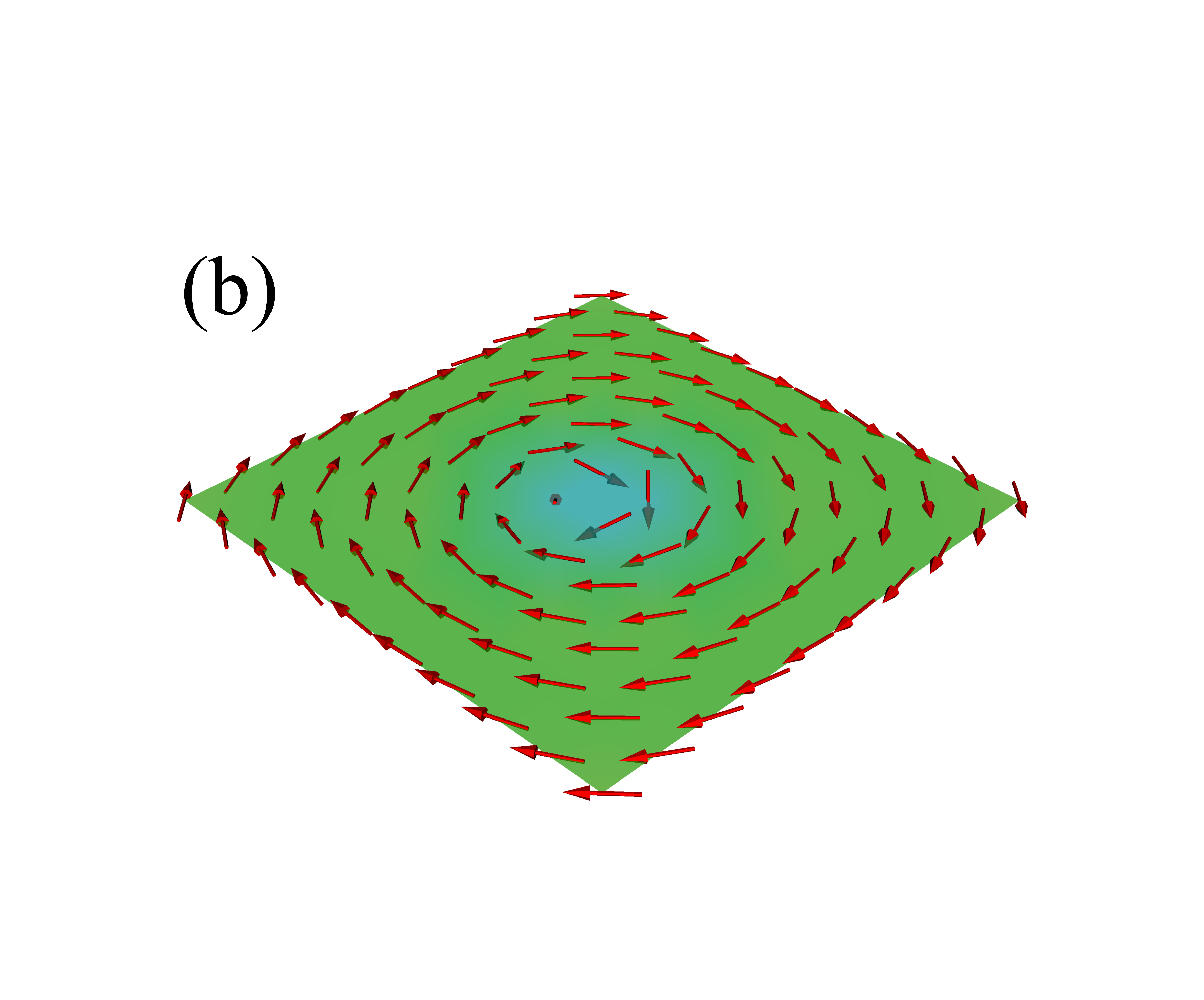}
\includegraphics[width=4cm]{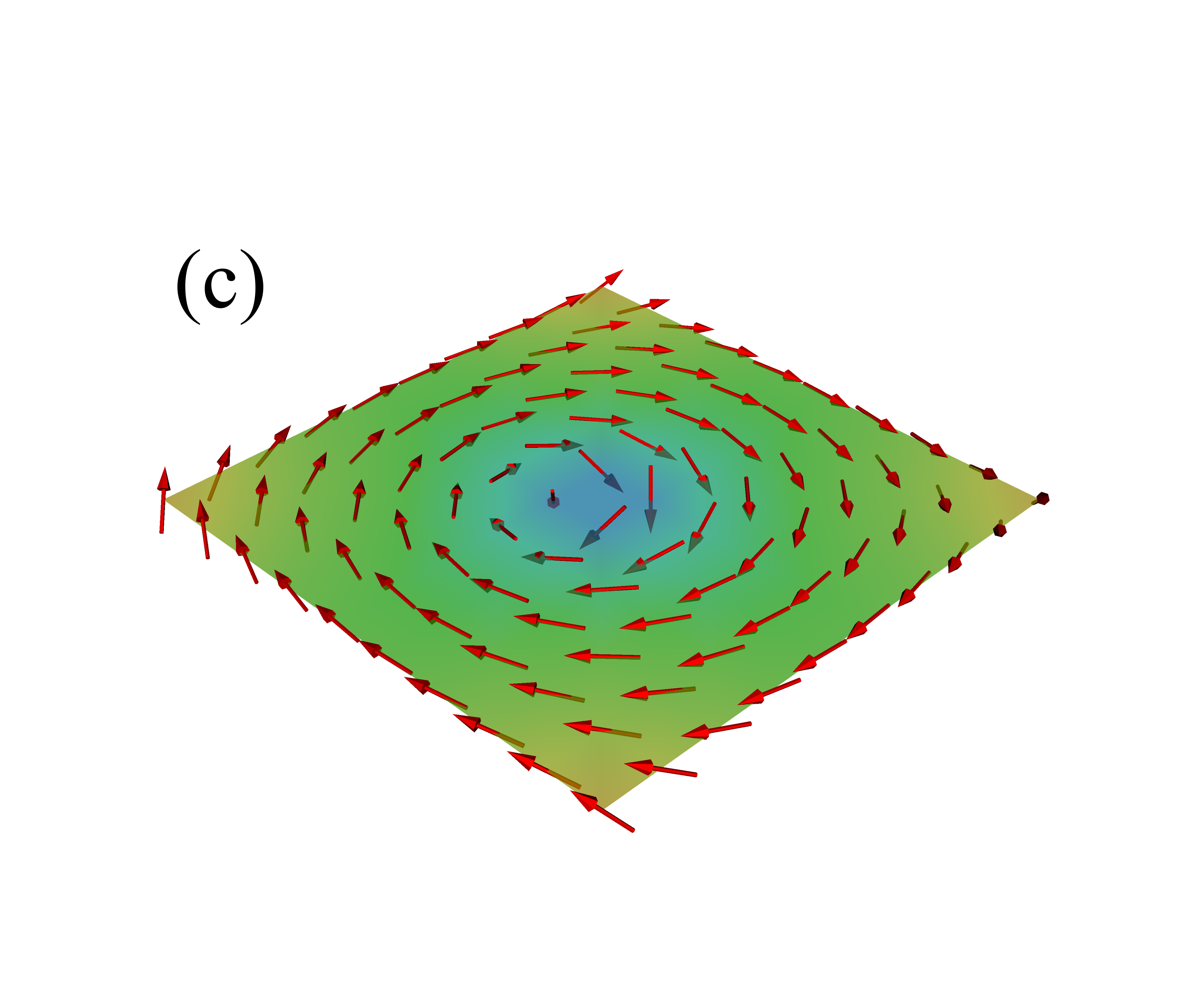}
\includegraphics[width=4cm]{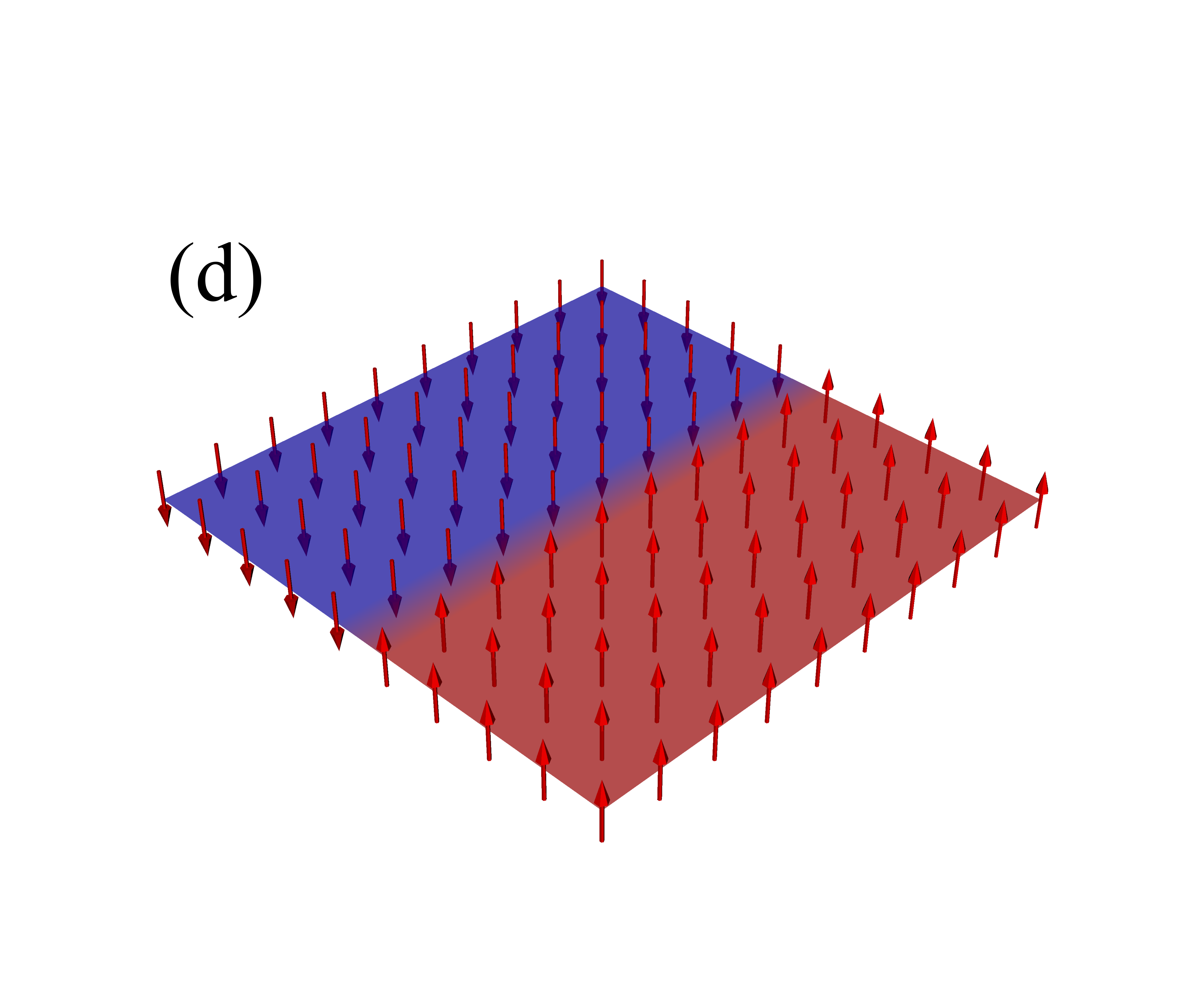}
	\caption{(Color online) Ground state structures established for the below-specified ranges of perpendicular anisotropy (parameter $A$), with the following values of the other parameters assumed: nanodot size $L = 10$, exchange integral $J = 1$ and dipolar coupling $D = 0.3$.  The colors indicate the orientation of spins: green for in-plane spins, blue for down spins and red for up spins.  The intensity of each color expresses the degree of spin aligning: (a) planar vortex $0<A<0.75$, (b) vortex with core $0.75<A<0.95$, (c) vortex with core and corners $0.95<A<1.57$, (d) domain structure (stripes) $A>1.57$.}
	\label{fig:Pic0o3}
\end{figure}

To investigate the evolution of the GS with growing anisotropy let us first consider the case shown in Fig.~\ref{fig:Pic0o3}, with $L=10$ and the other parameters fixed at $J=1$ and $D=0.3$. The colors indicate the spin orientations: green for in-plane, blue for down, and red for up spins. When the anisotropy is small, the GS is seen to be a planar vortex state, with all the spins lying in the plane of the system. However, above a certain threshold anisotropy value (in this case $A=0.75$) the spins begin to draw forward from the plane of the system in its central region, which we shall refer to as the vortex core; both magnetization directions - upward and downward - are energetically equivalent. As $A$ continues to grow, above a second threshold value (of 1.57) the minimum energy configuration becomes a stripe structure with alternating stripes of upward and downward spins perpendicular to the plane of the system. (It is noteworthy that the threshold value of 0.75 for which the core has emerged in our study is very close to that calculated analytically \cite{Wysin49} and numerically \cite{Gouvea39} for a spin system of similar size, but only with the exchange interaction.)

%
\begin{figure}[h]
	\centering
	\includegraphics[width=4cm]{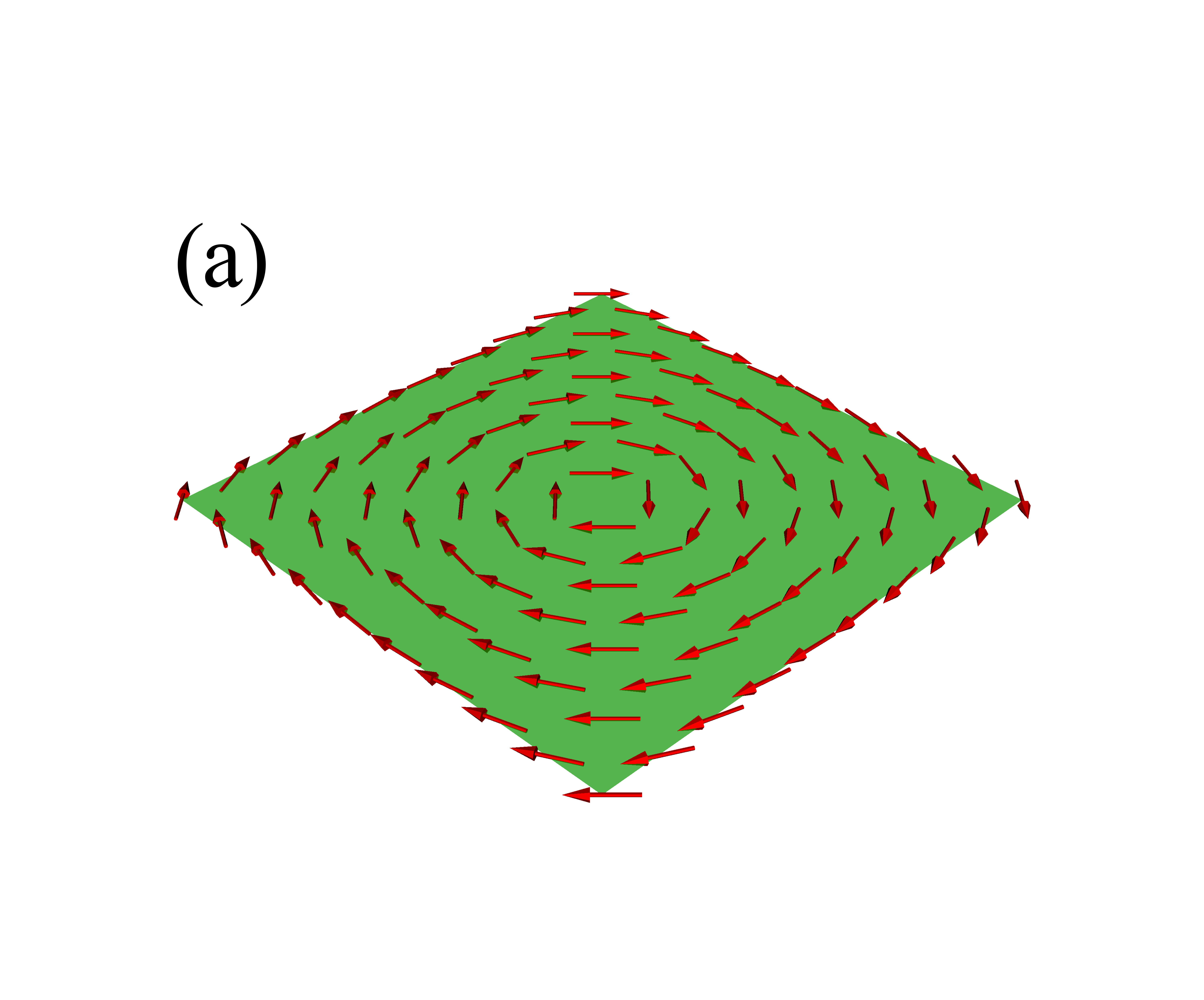}
\includegraphics[width=4cm]{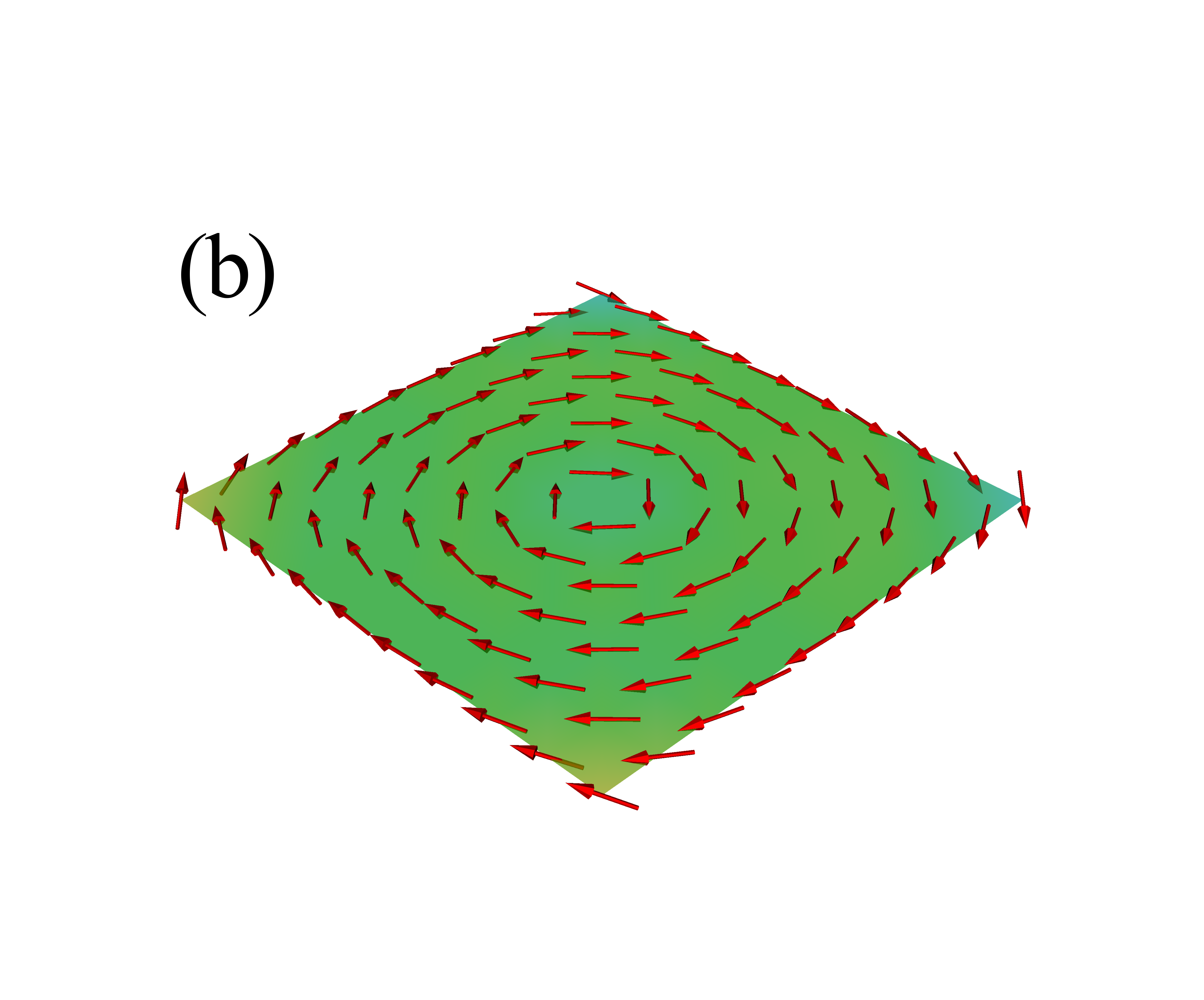}
\includegraphics[width=4cm]{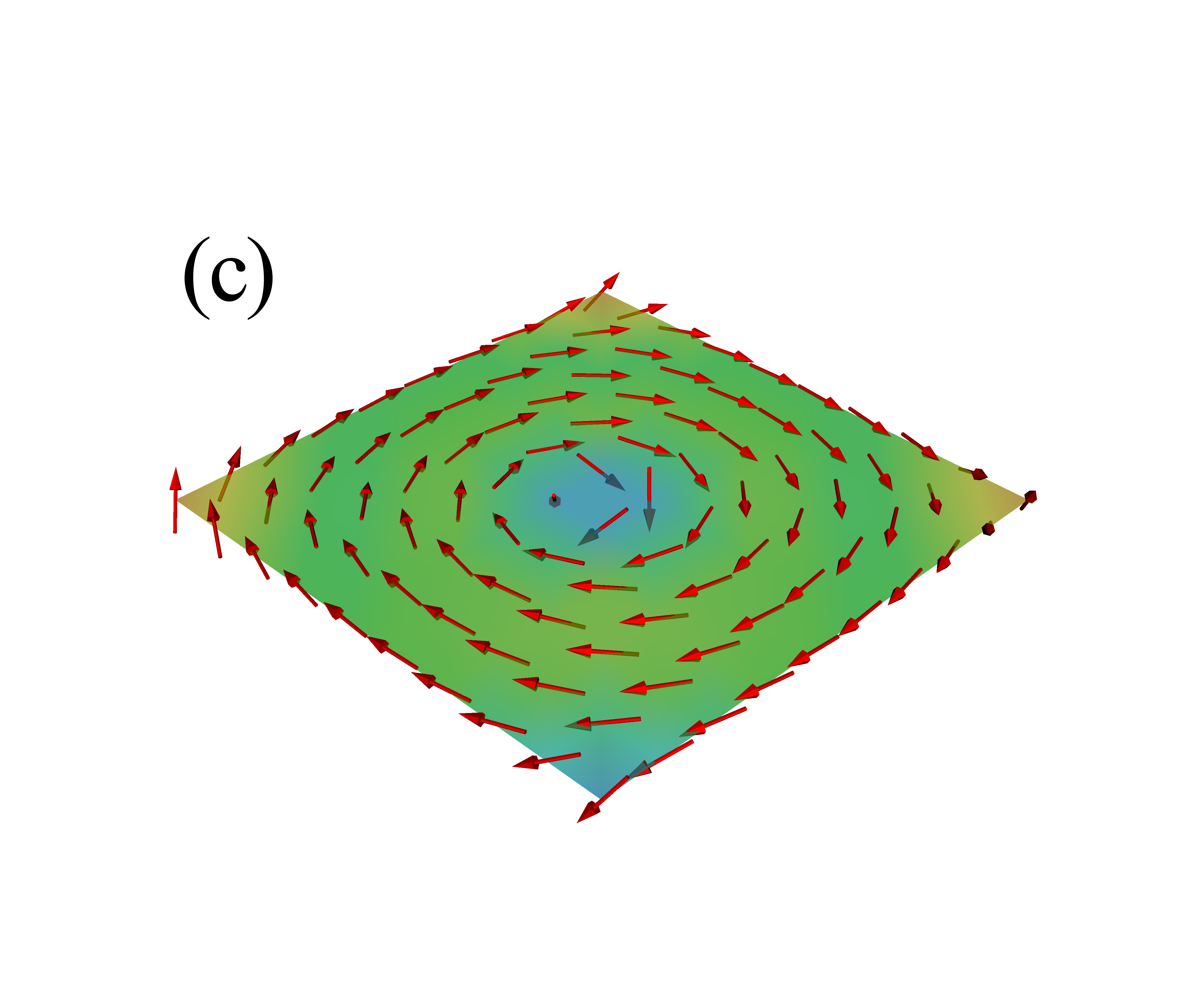}
\includegraphics[width=4cm,angle=0]{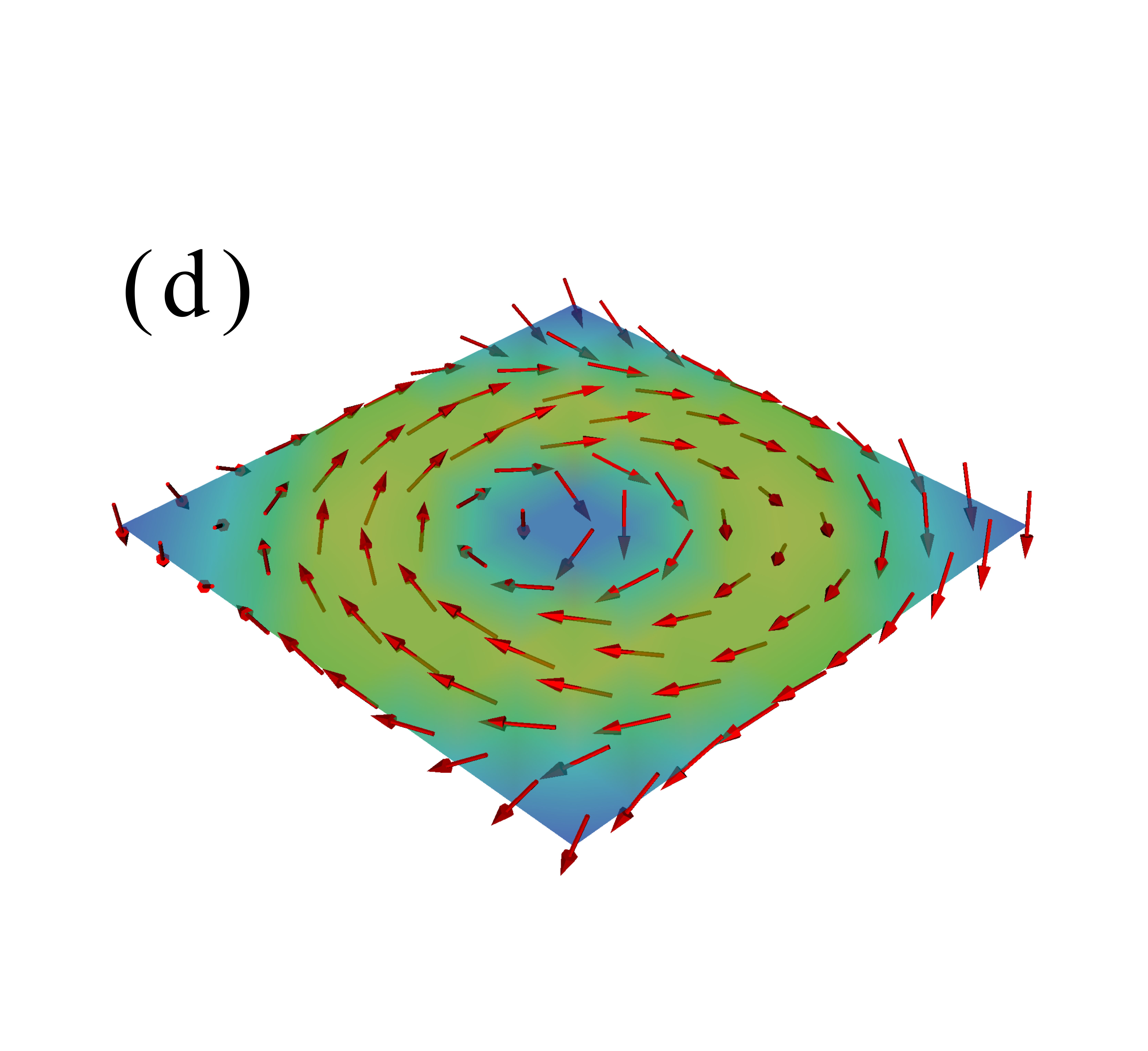}
\includegraphics[width=4cm,angle=0]{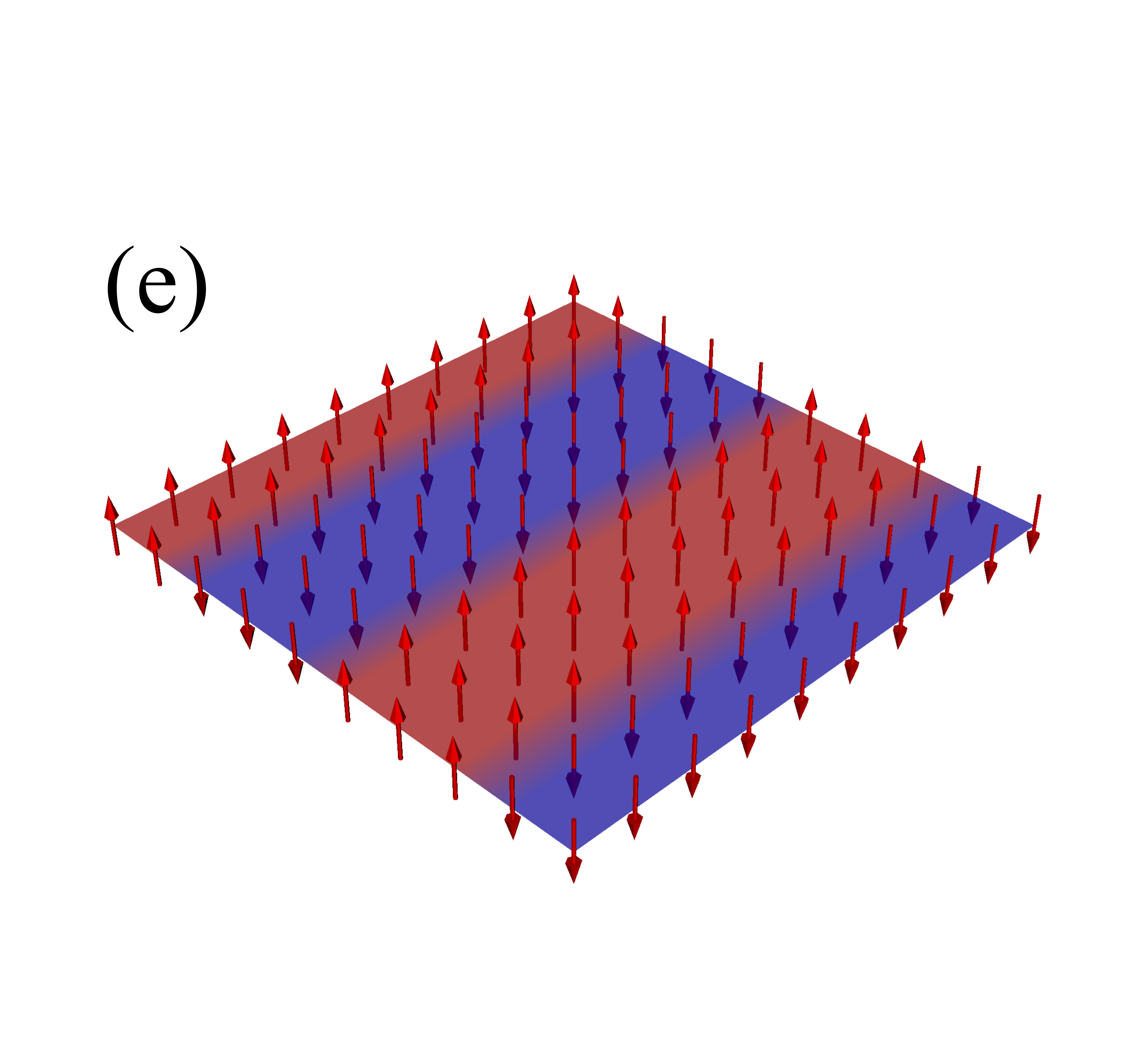}
	\caption{(Color online) Ground state structures established for the below-specified ranges of perpendicular anisotropy (parameter $A$), with the following values of the other parameters assumed: nanodot size $L = 10$, exchange integral $J = 1$ and dipolar coupling $D = 0.5$.  The same color codes as in Fig. 1 are used: (a) planar vortex $0<A<1.32$, (b) vortex with corners $1.32<A<1.47$, (c) vortex with core and corners $1.47<A<1.87$, (d) satellite formation $1.87<A<2.37$, (e) domain structure (stripes) $A>2.37$.}
	\label{fig:Pic0o5}
\end{figure}
Now, let us scrutinize the evolution, shown in Fig.~\ref{fig:Pic0o5}, of the GS of the system with much stronger dipolar interaction ($J=1$, $D=0.5$). The nonzero magnetization regions are seen to start forming at the corners rather than in the center of the sample (the threshold value of $A$ is 1.32); only above a second threshold value, $A=1.47$, does the core emerge. The next configuration, which occurs for $A$ greater than 1.87, is characterized by satellite regions of perpendicular magnetization around the core; the magnetization polarization in these regions is opposite to that in the core. Finally, in its further evolution the system achieves a stripe domain structure, which forms above another threshold value, 2.37, of the anisotropy parameter $A$.
%
%
\subsection{Stripe structures}
\label{sec:StrukturyPaskowe}
\begin{figure}[ht]
\centering
		\includegraphics[width=7cm]{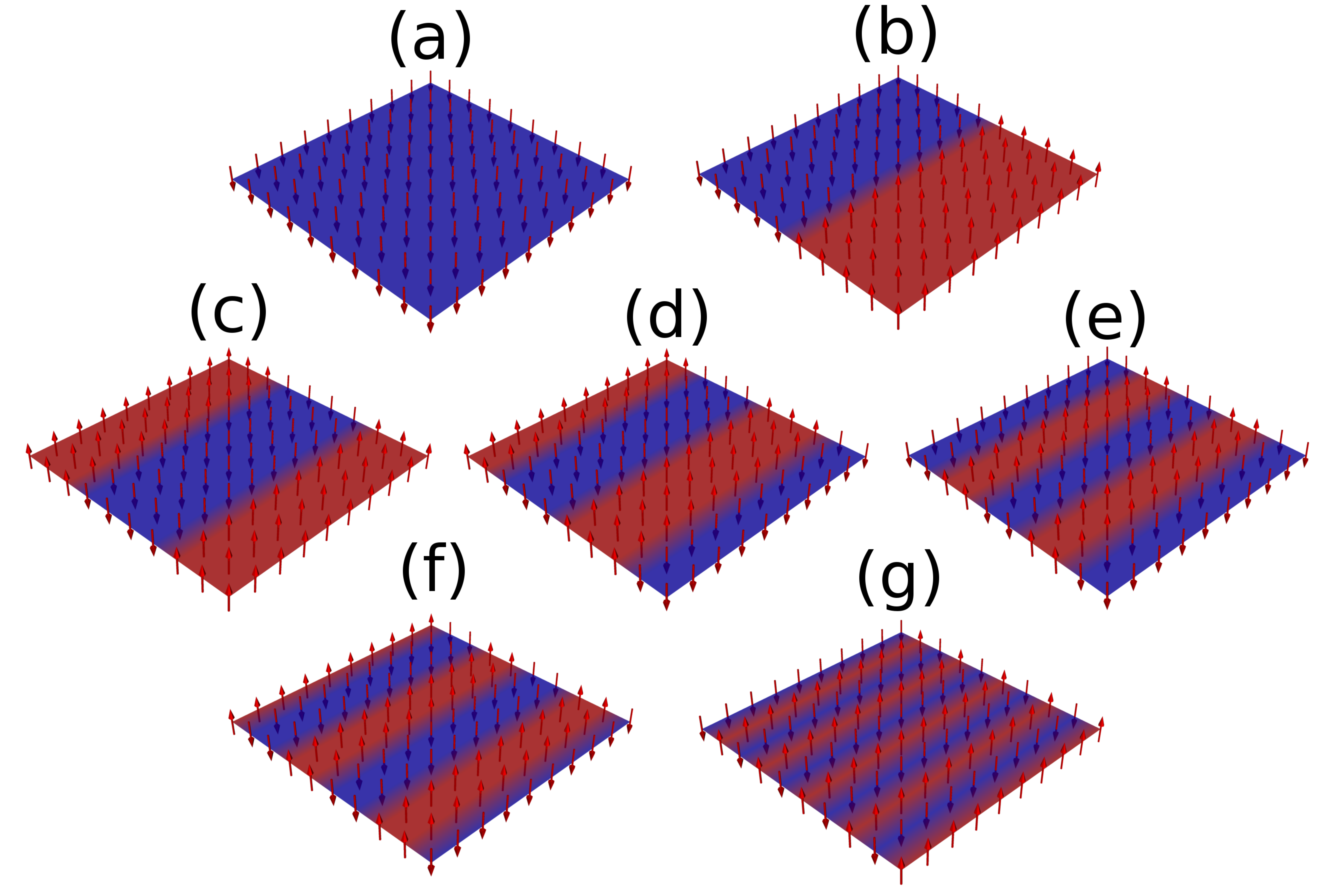}
	\caption{(Color online) Stripe domain ground-state structures established for different values of the $D/J$ in the giant perpendicular anisotropy regime: (a) $D/J$=0.00-0.25,
(b) $D/J$=0.25-0.32, (c) $D/J$=0.32-0.40, (d) $D/J$=0.40-0.50, (e) $D/J$=0.50-0.58, (f) $D/J$=0.58-0.83, (g) $D/J$=0.83-4.   Simulations were performed for $L=10$, $J=1$ and $A=4$. The same color codes as in Fig. 1 are used.}
	\label{fig:Stripes}
\end{figure}

\begin{figure}[htbp]
	\centering
		\includegraphics[width=8cm]{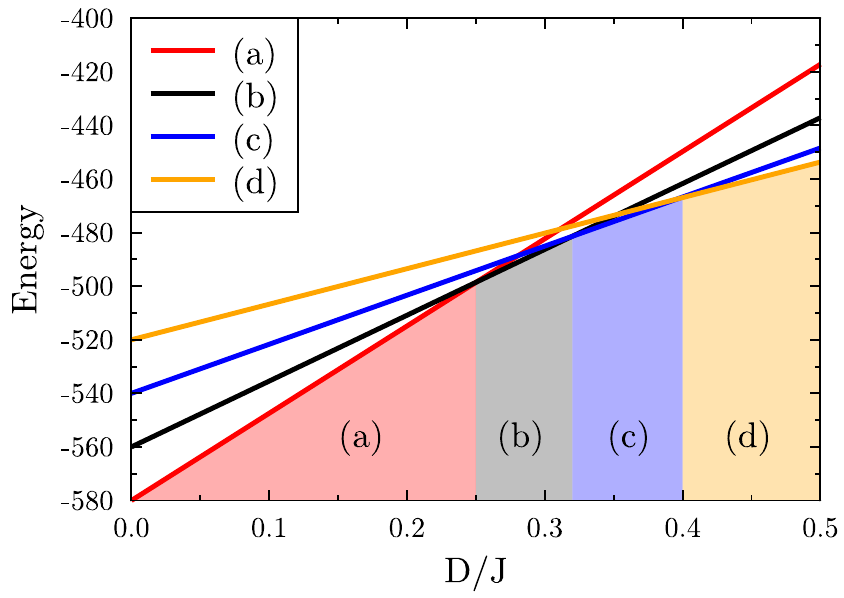}
	\caption{(Color online) Total energy (computed from Eq.~(\ref{eq:ham2})) of the spin structures shown in Fig.~\ref{fig:Stripes} vs. $D/J$. These plots provide a basis for determining the minimum-energy structure for a given value of $D/J$.}
	\label{fig:EnergyOfStripes}
\end{figure}

Thus, our simulations indicate that the stripe structure is the final configuration which the spin system seeks in the giant perpendicular anisotropy regime. This finding is easy to accept as with such a strong anisotropy all the spins are drawn out of the sample plane to set along its normal, i.e. $\vec{s}_i=(0,0,\pm1)\text{ for all } i$. Then the Hamiltonian (\ref{eq:ham}) takes the form:
\begin{equation}\label{eq:ham2}
 H= -J\sum_{ij}^{\text{nn}} s^{z}_is^{z}_j + D\sum_{ij}^{\text{all}} \frac{s^{z}_is^{z}_j}{r_{ij}^3} - AL^2,
\end{equation}
which implies that the anisotropy term only shifts the system energy level; for this very reason the number of stripe domains (and their pattern) at a given GS proves to  depend only on the ratio $D/J$, and not on $A$. Referring to the Hamiltonian (\ref{eq:ham2}), we can precisely indicate the ranges in which the stripe structure minimizes the system energy. The results of our investigations are shown in Fig.~\ref{fig:Stripes}, presenting the evolution of the stripe system with growing $D/J$ ratio. In the figure caption, we specify the stability ranges for each of the depicted stripe structures. The way in which these ranges have been determined is made clear in Fig.~\ref{fig:EnergyOfStripes}, which allows to compare the total energy of the considered stripe structures (for a specific $D/J$ ratio) to find the lowest-energy structure.

Let us briefly summarize and give a qualitative explanation on the GS's found above. For very low $A$, the dipolar term dominates, yielding an in-plane vortex configuration. For larger $A$, spins are ordered in vortex configurations with out-of-plane core and corner spins which is a compromise between $D$ and $A$ effects.  For very large $A$, the energy is lowest when spins are perpendicular. In this situation, the effects of $J$ and $D$ come to determine whether they should be up or down or both:  $J$ favors parallel neighbors while $D$ favors antiparallel spins, as seen by examining the signs of these terms in Eq. (\ref{eq:ham2}).  Increasing $D$ will thus favor antiparallel ``domains". Therefore, the larger $D$ becomes the more numerous domains are created, as we observe in Fig. \ref{fig:Stripes}.

%
%

\section{Finite-temperature properties}\label{FTB}
In statistical physics, a phase transition is defined for systems at the thermodynamic limit where thermodynamic functions diverge or undergo anomalies. Two often encountered types of phase transition are (i) the second-order transition where the second derivatives of the free energy, such as the specific heat and the susceptibility, diverge at the infinite system size, (ii) the first-order transition where the first derivatives of the free energy, such as the internal energy and the magnetization, have  discontinuity.  We cannot theoretically define a phase transition for a finite-size system.  In simulations and in theories, we can study finite-size systems but we use the finite-size scaling\cite{Landau09,Zinn,Amit} to predict the characteristics of the phase transition at the infinite-size limit.  In nanodots with short-range interaction for instance, we cannot talk about phase transition in the sense of the above definition: due to their very small sizes, the spins spend a finite time to reverse their orientation over and over again during the simulation time so that there is no stabilized ordering.   However,  in systems with long-range interaction where each spin is in interaction with all others as the dipolar case treated here,  the ground-state ordering is due to the whole all-spin connection: exciting a spin costs an important energy amount, unlike in the short-range interaction case. As it turns out, the GS found above is stabilized at finite temperatures and is destroyed only at a higher temperature. We shall use the term ``transition" to indicate this change of ordering which is seen by the variation of the order parameter, even if this transition does
not occur with a divergence of physical quantities since we work with nanodots.  The dependence
of physical quantities on the system size in the transition temperature region proves that the anomaly is
indeed a phase transition if we let the system size go to infinity.  We will return to this point below.

A second point which is important to emphasize is the following.  We know from the Mermin-Wagner theorem \cite{Mermin}  that systems of spins of continuous degrees of freedom, such as XY or Heisenberg spins, with isotropic short-range interactions do not have long-range ordering at finite $T$ in 2D. Our present model, though having a very small size, possesses the long-range interaction and an Ising-like anisotropy, namely the two factors which favor the ordering at finite $T$.
We will see below that we have, even in the most unfavorable case with small $A$, a sharp peak of the specific heat at the loss of vortex ordering at a finite $T$.

We use the standard MC simulation method which is enough for our purpose.  The histogram method is used to detect first-order transition when necessary. In general, we discard about $10^6$ MC steps per spin for equilibrating and average physical quantities over the next $N_{MC}=10^6$ MC steps per spin.  We will concentrate ourselves in the following to the case $D=0.3$ with varying $A$,  for numerical presentation.  Note however that physical behaviors depend roughly on the ratio $A/D$.

Let us define the following order parameters depending on the phase symmetry:

(i) In-plane vortex phase:
\begin{eqnarray}
M_v&=&\frac{1}{N\times N_{MC}}\nonumber\\
&&\times \sum_{t=1}^{N_{MC}} \left |
 \left [ \sum_{i}^{\text{all}}[\vec u_i\wedge \vec s_i(t)]/\sin a_0\right]_z\right |\nonumber\\
 &&
\end{eqnarray}
where $N=L^2$, $\vec u_i=\vec r_i/r_i$ (unit vector along the vector $\vec r_i$ connecting the center of the dot and the lattice site $i$), $a_{0}$ is the angle between the spin $\vec s_i$ and $\vec u_i$ in the GS.
Note that $M_v$ is used only when the GS is not ferromagnetic.
At $T=0$, if the configuration is planar, we have $[\vec u_i\wedge \vec s_i/\sin a_0]_z=1$ for any $i$, so that $M_v=1$.  If the configuration is perpendicular, $[\vec u_i\wedge \vec s_i]_z=0$ for any $i$, so that  $M_v=0$.  When the system is disordered $M_v$ =0 because  $\vec u_i\wedge \vec s_i(t)$ is a random vector.  In the vortex GS configuration with core, spins at the dot center are not in the $xy$ plane, so the value of $[\vec u_i\wedge \vec s_i]_z$ is between 0 (if perpendicular ) and $q< 1$ (if tilted). Hence $M_v$ is not saturated at 1 in the GS as seen below.

(ii) Uniform perpendicular configuration:
\begin{equation}
M_z=\frac{1}{N \times N_{MC}}\sum_{t=1}^{N_{MC}} \left | \sum_i^{\text{all}} s_i^z(t)\right |
\end{equation}

(iii) Stripe configuration:
\begin{equation}
M_s=\frac{1}{N\times N_{MC}}\sum_{t=1}^{N_{MC}}  \left |\sum_i^{\text{all}} (-1)^ps_i^z(t)\right |
\end{equation}
where $p$ is the ``parity" of the $z$ domain ($p=\pm 1$ for GS down and up domains respectively).

\begin{figure}[htbp]
	\centering
\includegraphics[width=5.5cm,angle=0]{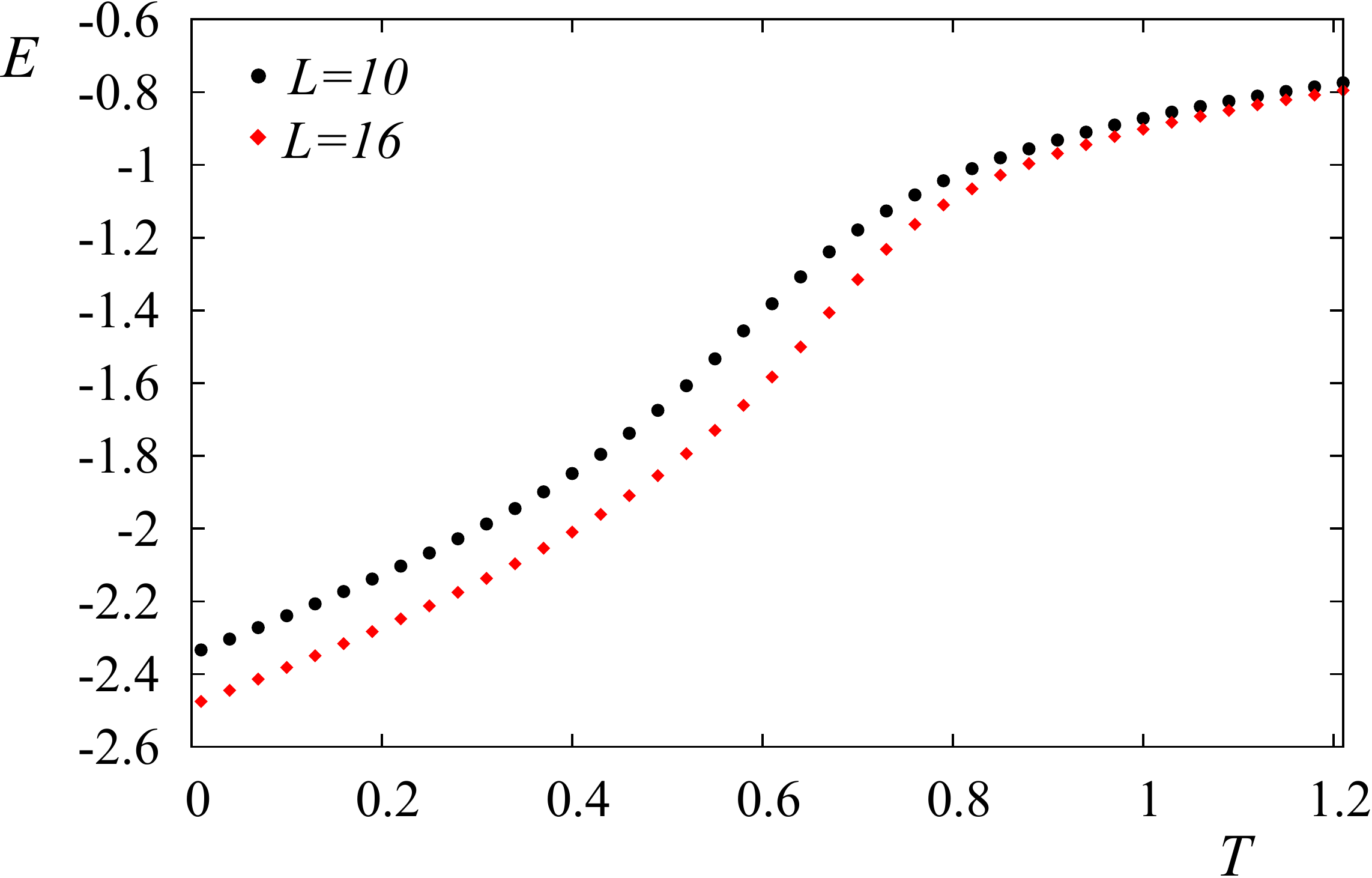}
\includegraphics[width=5.5cm,angle=0]{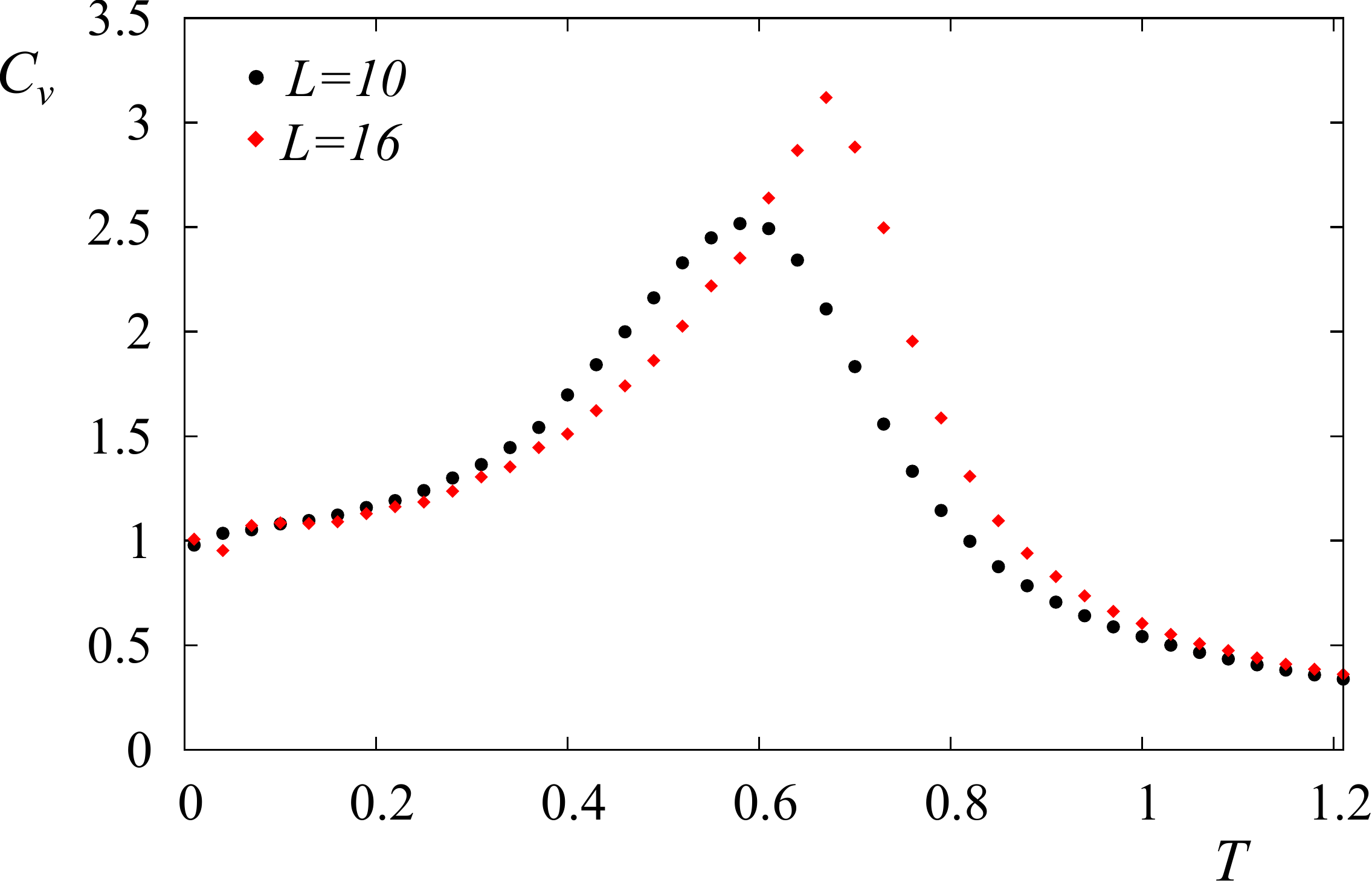}
\includegraphics[width=5.5cm,angle=0]{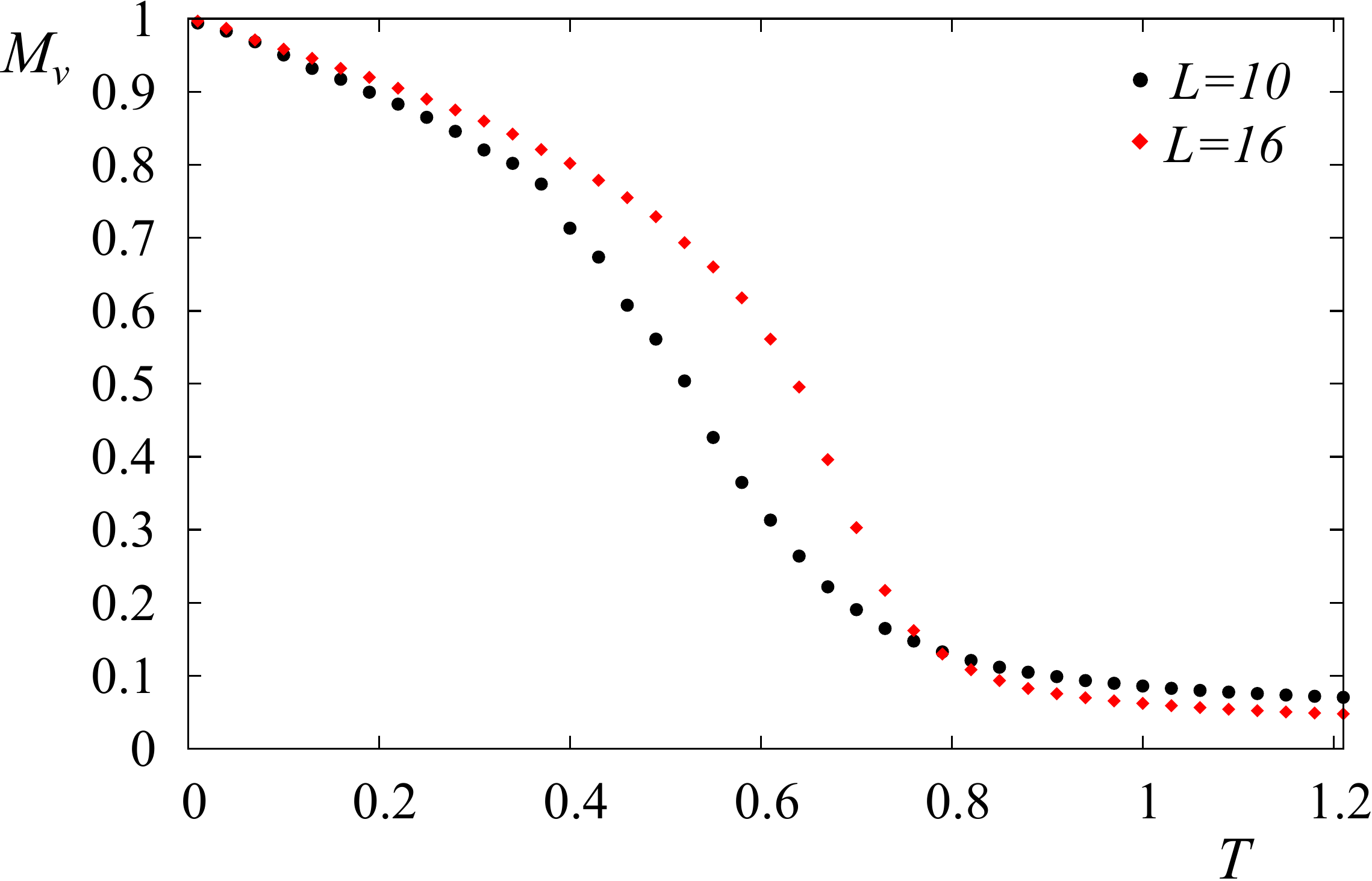}
	\caption{(Color online) Top: Energy per spin, Middle: Specific heat, Bottom:  In-plane
vortex order parameter $M_v$,  versus $T$ for $A/D=2$ with $L=10$ (black circles) and $L=16$ (red diamonds). }
	\label{fig:ECM06}
\end{figure}

In the case of weak anisotropy, the dipolar interaction yields an in-plane vortex configuration without core, as seen above.  As $T$ increases, we observe a sharp transition from the vortex configuration to the disordered phase, as shown in Fig. \ref{fig:ECM06}, for $A/D=2$.  The same behavior is seen for $A/D=0$ to $\simeq 3$ for $L=10$.  When the dot size is increased, the dipolar contribution to the energy is larger, so the overall energy is more important as seen in  the figure at low $T$, making the transition temperature higher.  Note that the peak of the specific heat $C_v$ is very sensitive to the system size, indicating that it corresponds to a real phase transition.  The same effect is seen in the order parameter $M_v$ shown in
Fig. \ref{fig:ECM06}: increasing $L$ makes the fall of $M_v$ much sharper at the transition temperature, namely a higher peak of the susceptibility (not shown). It is interesting to study systematically many larger sizes and to use the finite-size scaling \cite{Landau09,Zinn,Amit,Privman,DiepTM} to determine the type of the transition, namely its universality class. Such a formidable task is out of the scope of this paper, it is left for a future investigation.

\begin{figure}[htbp]
	\centering
\includegraphics[width=5.5cm,angle=0]{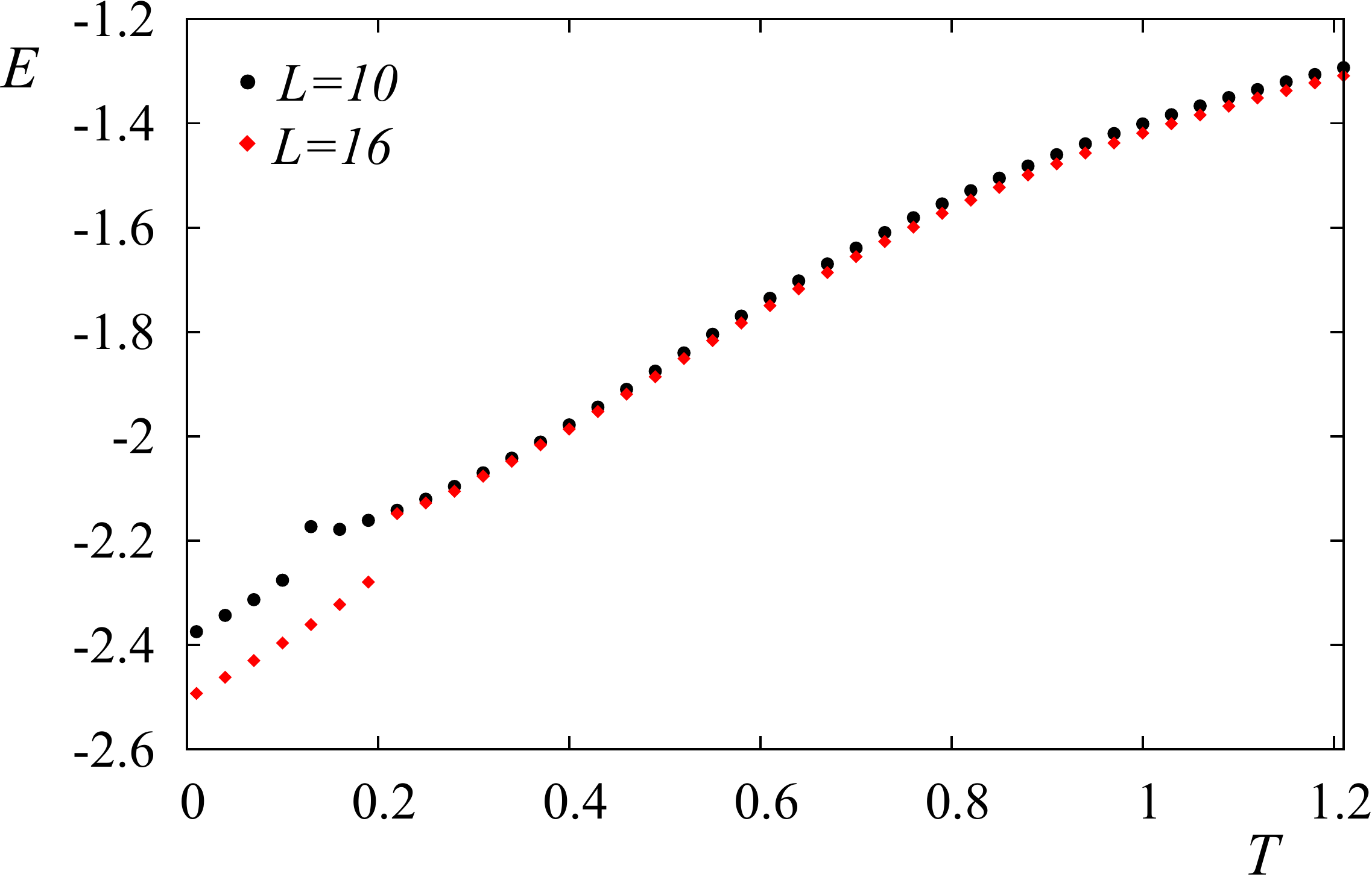}
\includegraphics[width=5.5cm,angle=0]{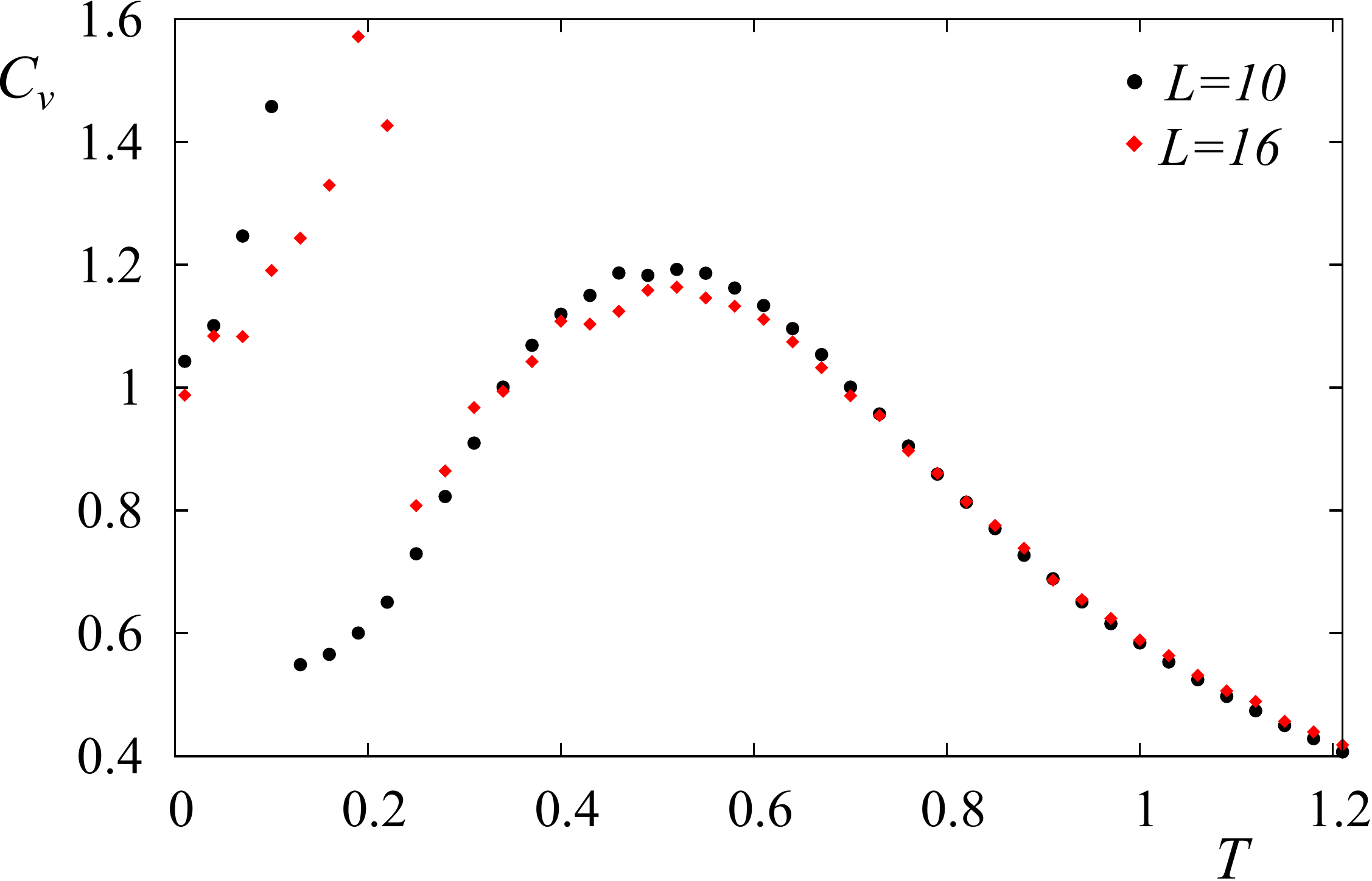}
\includegraphics[width=5.5cm,angle=0]{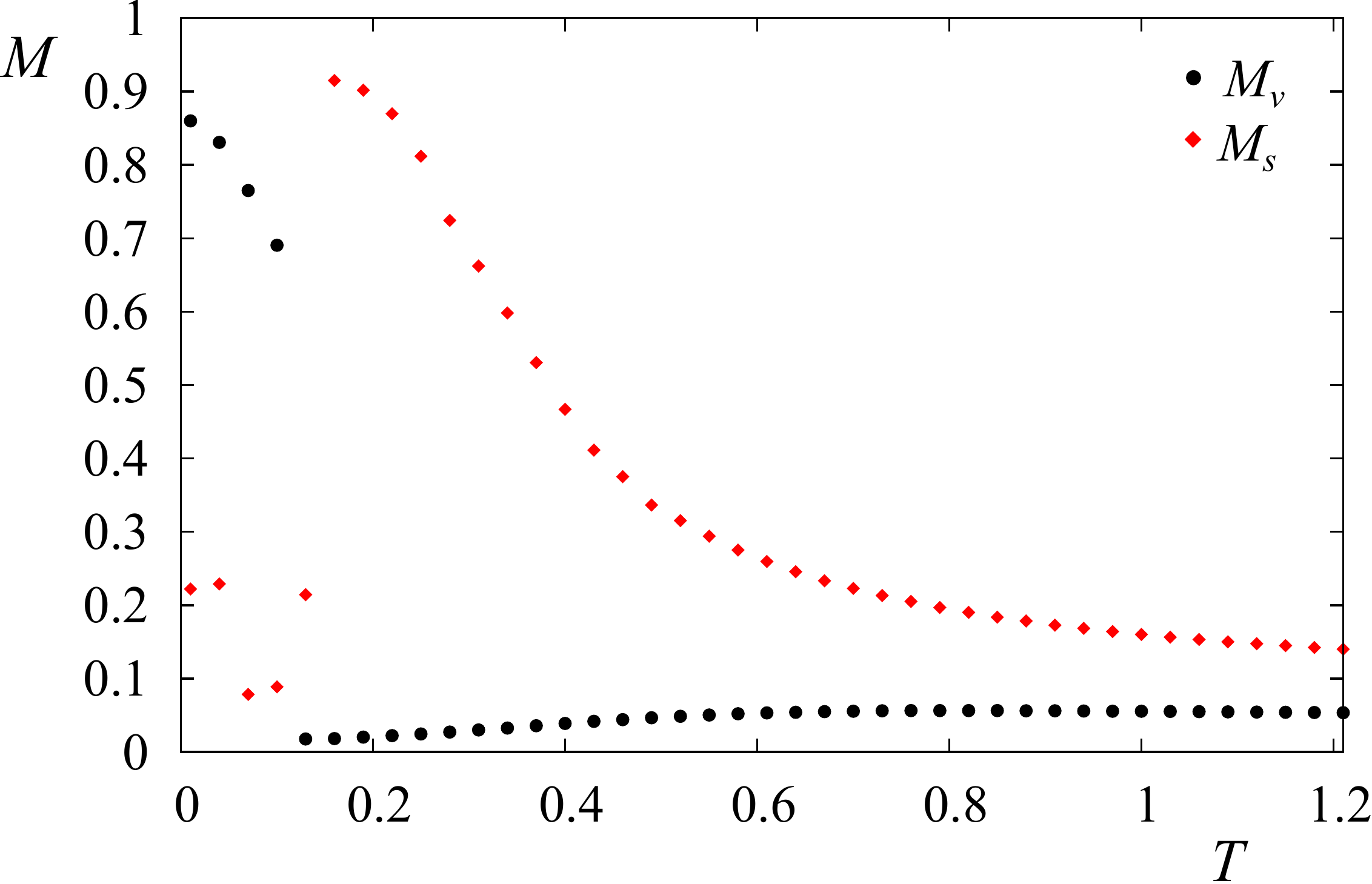}
	\caption{(Color online) Top: Energy, Middle: Specific heat, Bottom:  In-plane
order parameter $M_v$ and perpendicular stripe magnetization $M_s$, versus $T$, for $A/D=1.4/0.3\simeq 4.6$, with $L=10$ (black circles) and $L=16$ (red diamonds).  }
	\label{fig:ECM14}
\end{figure}

For larger anisotropies,  there is a first-order transition at low $T$ as seen in Fig. \ref{fig:ECM14} with $A/D=1.4/0.3 \simeq 4.6$: this transition occurring at $T\simeq 0.1$ changes the system ordering from the vortex configuration with core to the perpendicular stripe configuration.   The energy undergoes a large discontinuity at the transition. It is interesting to note that the vortex core at $T=0$ is characterized by a non zero $M_z$ and a non saturated in-plane order parameter $M_v$.  $M_z$ decreases  when $T$ is increased from 0. At the transition from the in-plane vortex  to the perpendicular stripe configuration $M_s$ jumps to a high value while $M_v$ goes down to 0.  When $T$ is increased further, the perpendicular stripe  configuration becomes progressively disordered. The system is entirely disordered for  $T> 0.45$ for $L=10$.  This is not a phase transition because the peak of $C_v$ does not depend on $L$ as seen in Fig. \ref{fig:ECM14}.

\begin{figure}[htbp]
	\centering
\includegraphics[width=4cm,angle=0]{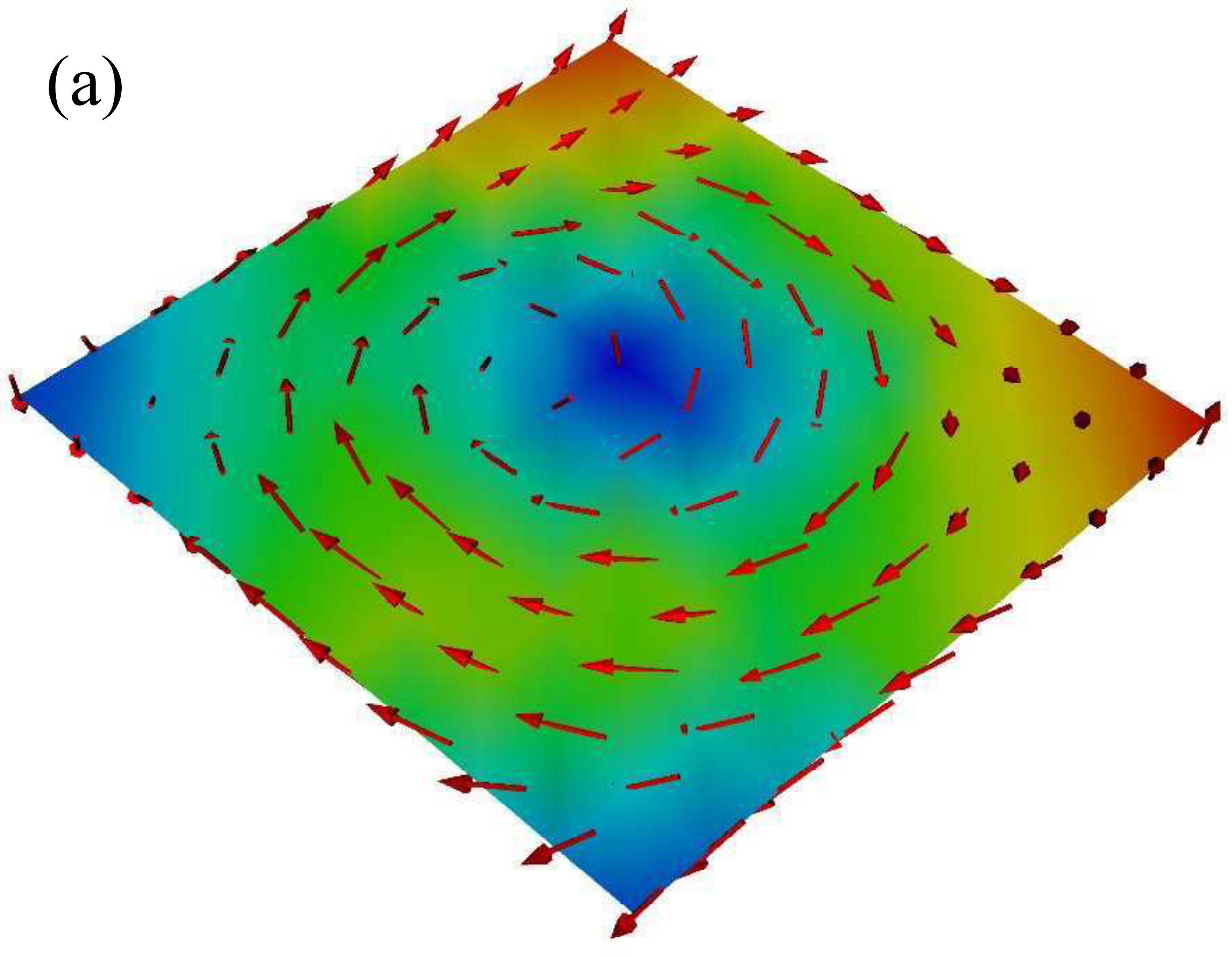}
\includegraphics[width=4cm,angle=0]{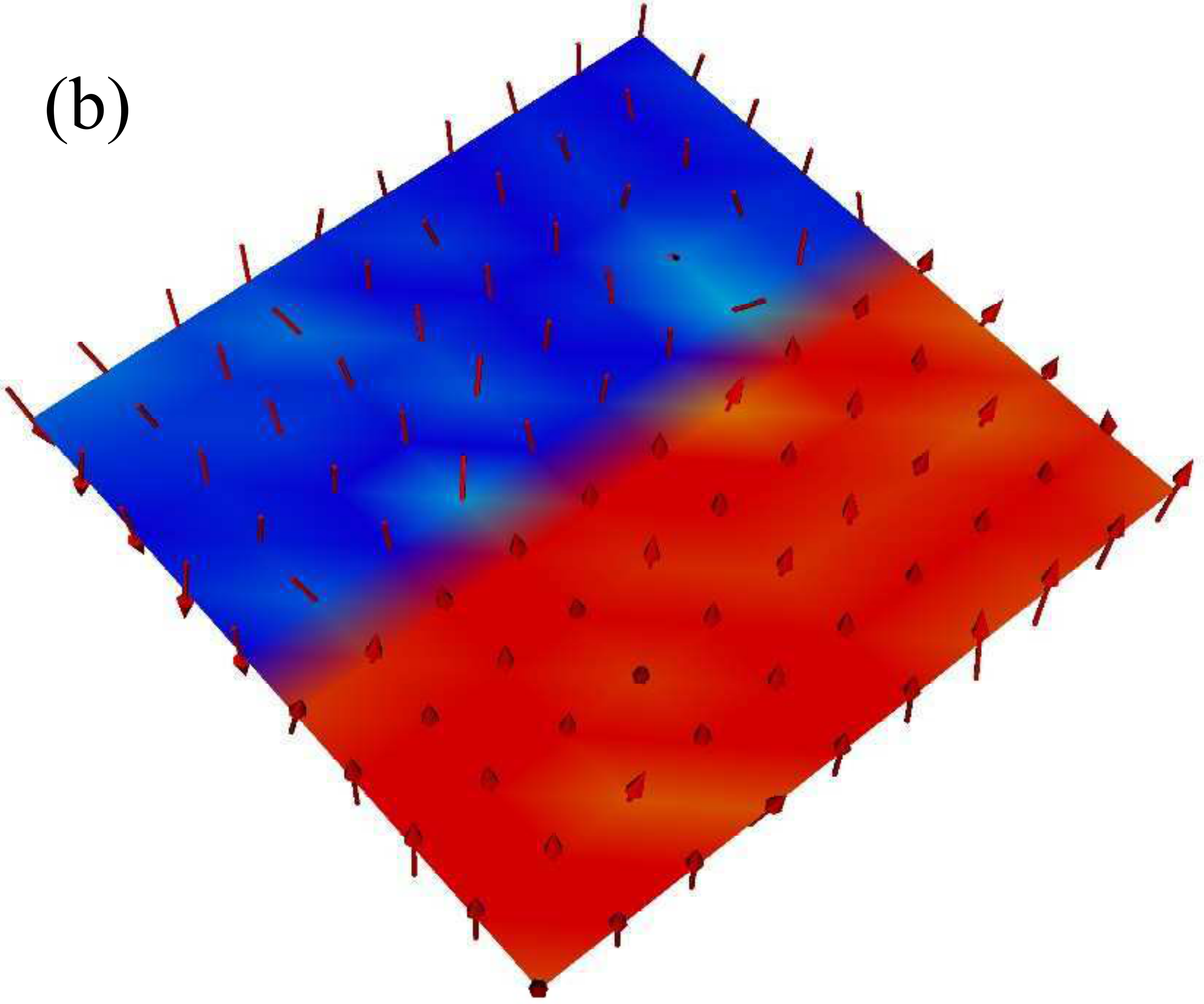}
\includegraphics[width=4cm,angle=0]{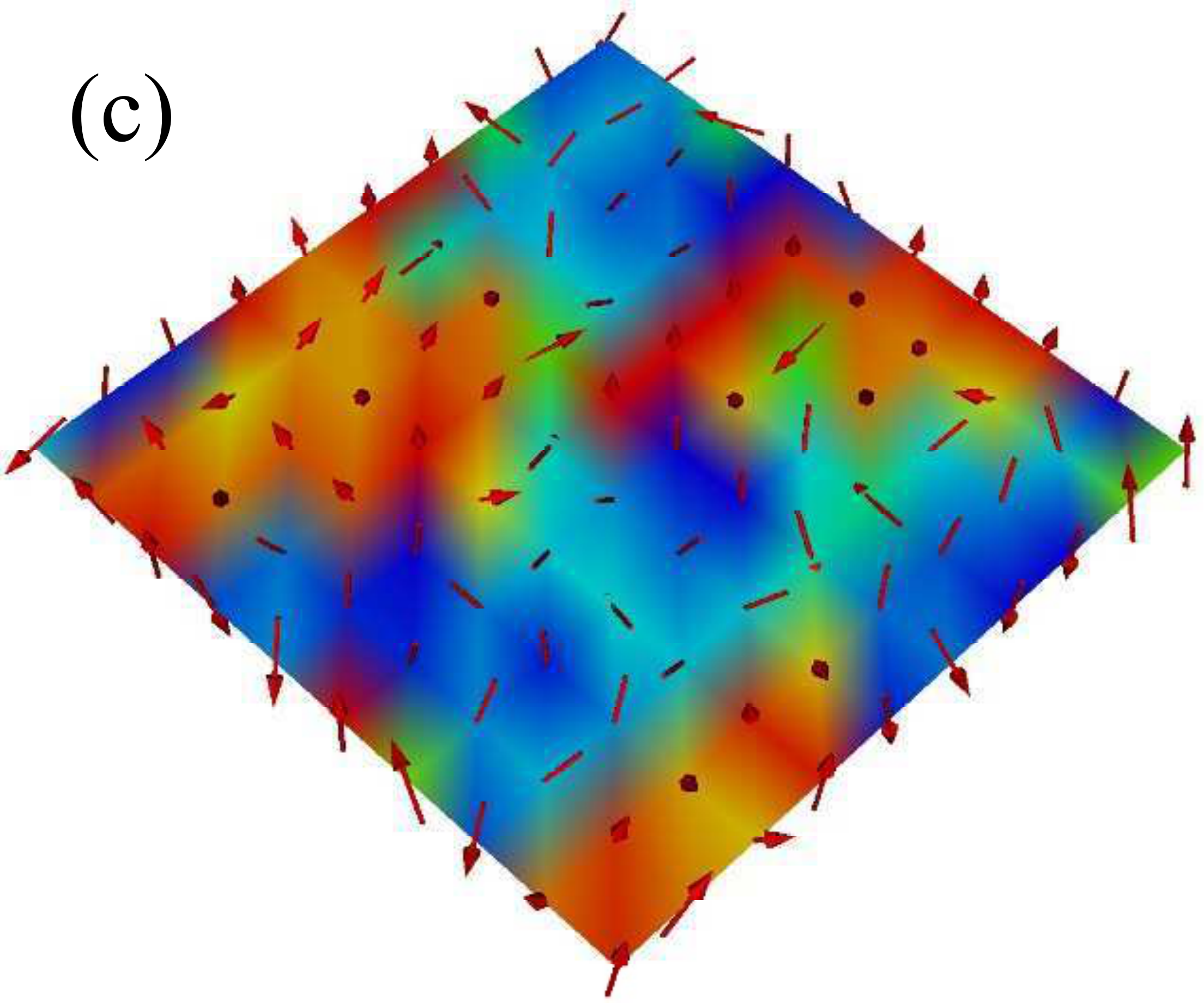}
	\caption{(Color online) Snapshots of the dot at (a) $T=0.01$, (b) $T=0.19$, (c) $T=0.82$,  for $A/D=1.4/0.3 \simeq 4.6$, $L=10$.  The same color codes as in Fig. 1 are used.}
	\label{fig:snapshot}
\end{figure}

We show in Fig. \ref{fig:snapshot} snapshots of the dot at three typical temperature regions: $T=0.01$ below the transition, $T=0.19$ in the perpendicular configuration, and $T=0.82$ in the disordered state.
We clearly see that the snapshot at $T=0.01$ shows the out-of-plane vortex at the core and at the corners.

\begin{figure}[htbp]
	\centering
\includegraphics[width=5.5cm,angle=0]{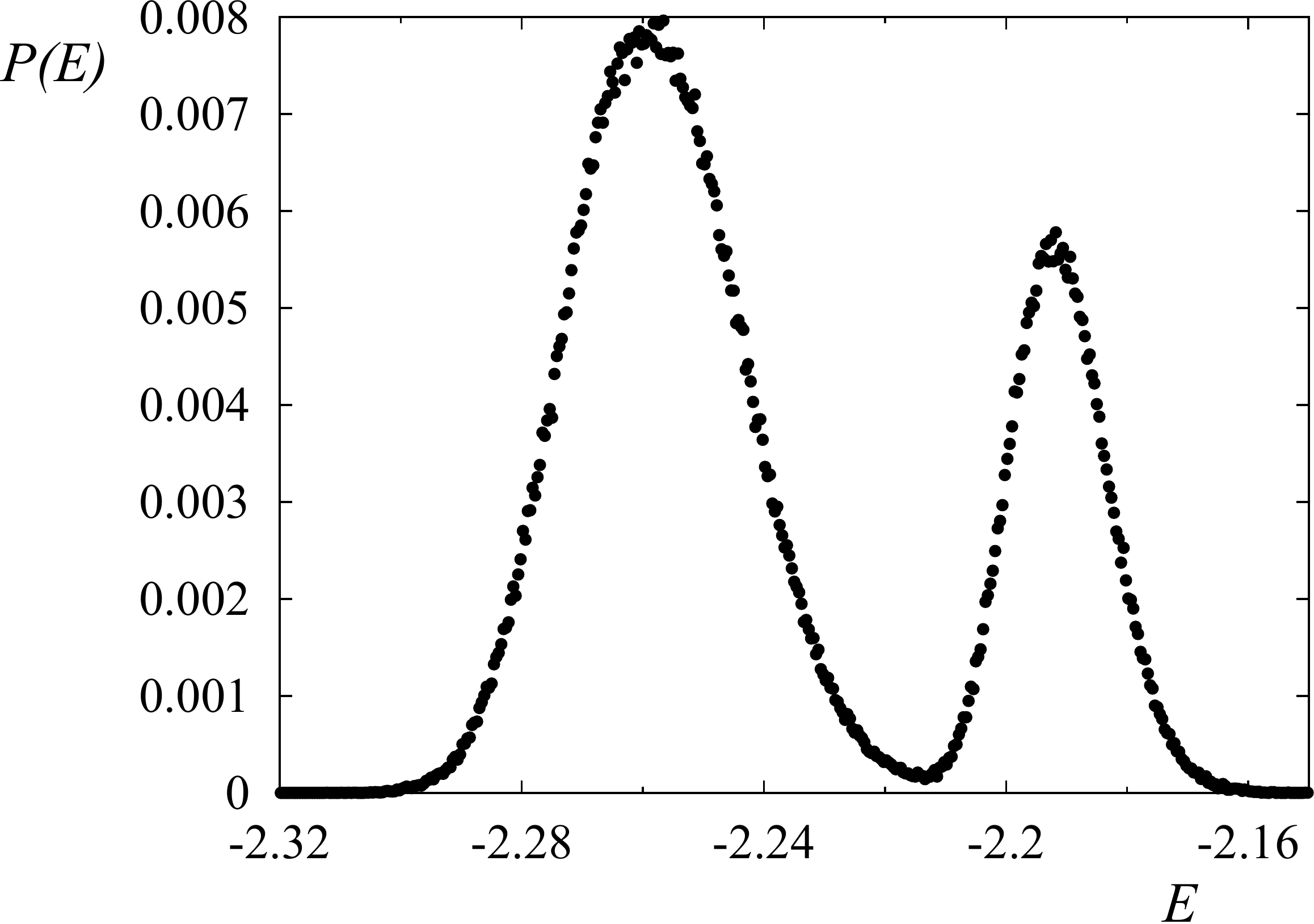}
	\caption{Energy histogram taken at the transition temperature $T_c=0.111$ of the case $L=10$ and $A/D=1.4/0.3$.}
	\label{fig:hist}
\end{figure}

The first-order transition is confirmed by the double peak structure in the energy histogram taken at the transition temperature $T_c=0.111$ shown in Fig. \ref{fig:hist}: the distance between the two peaks indicates a latent heat.

\begin{figure}[htbp]
	\centering
\includegraphics[width=5.5cm,angle=0]{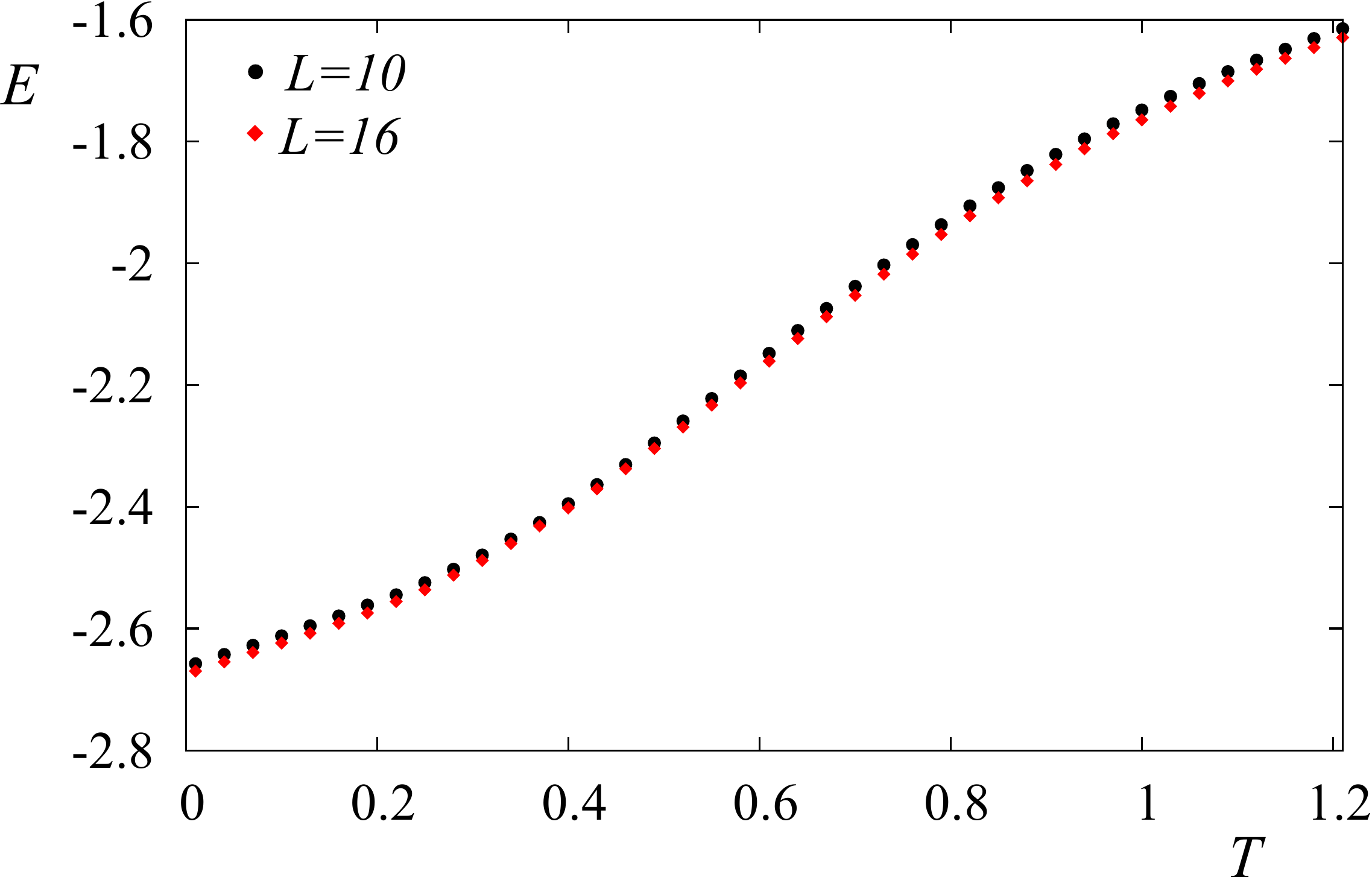}
\includegraphics[width=5.5cm,angle=0]{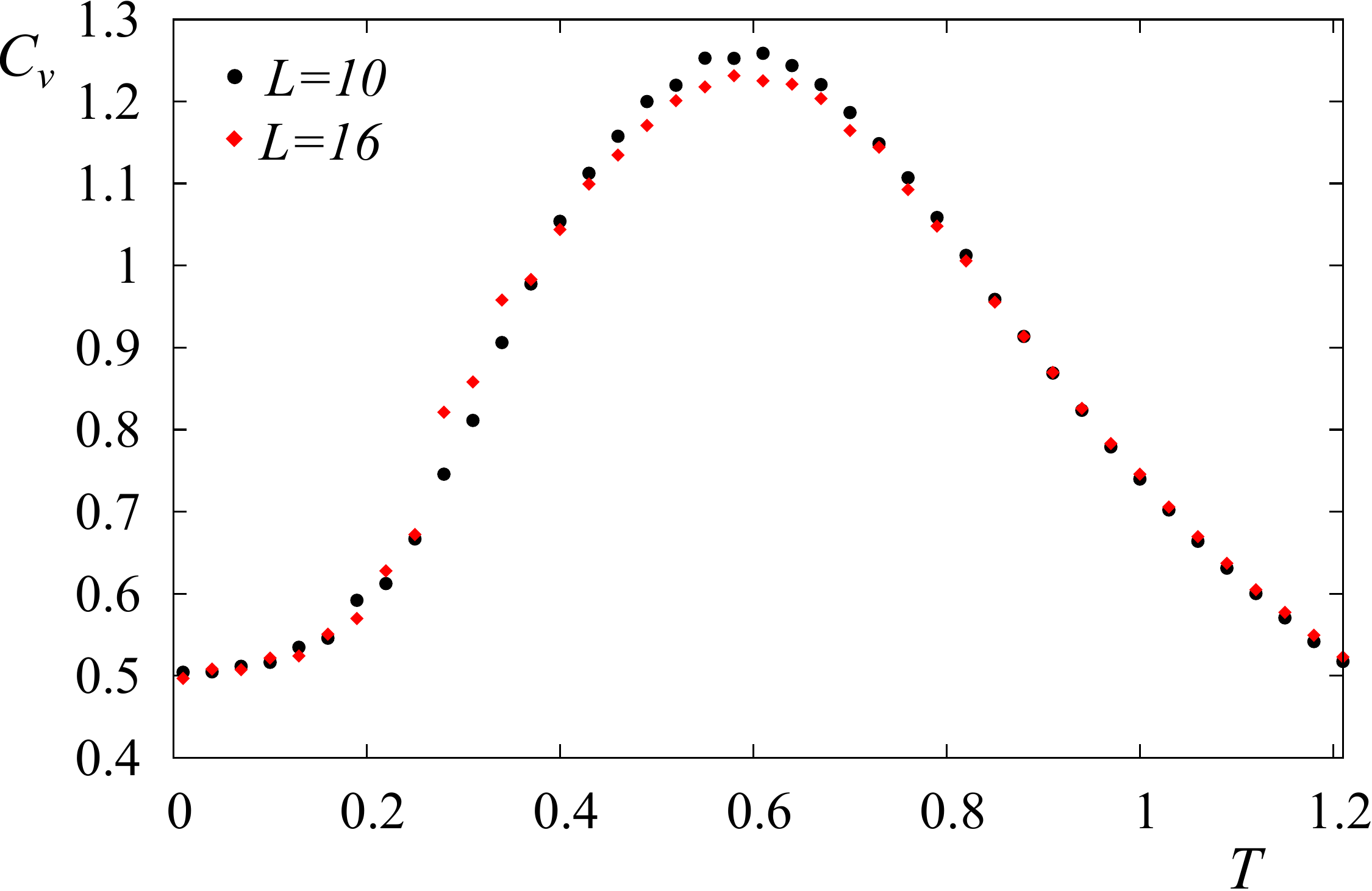}
	\caption{(Color online) Top:  Energy per spin, Bottom:  Specific heat,  versus $T$  for $A/D=1.8/0.3=6$, with $L=10$ (black circles) and $L=16$ (red diamonds). See text for comments.}
	\label{fig:ECM18}
\end{figure}

For very large anisotropies, namely for the GS stripe domains, there is no more phase transition as seen in Fig. \ref{fig:ECM18} where the peak of $C_v$ is constant with varying system size.  We emphasize here that when the spins are perpendicular to the plane, the first dipolar term in Eq. (\ref{eq:ham}) is zero (because $\vec{s}_i\cdot\vec{r}_{ij}=0$), and the second dipolar sum is very small due to the compensation of positive and negative energies of parallel and antiparallel spins  in the sum. Therefore,  for a larger dot size,  we have a larger dipolar sum but it does not change the energy per spin as seen in Fig. \ref{fig:ECM18}, unlike in the case of in-plane configuration shown in Fig. \ref{fig:ECM06} where the energy is lowered with increasing $L$,  resulting in a higher value of $T_c$.

\section{Conclusion}\label{Concl}
\label{sec:Conclusions}
We have used the simulated annealing MC method for investigating the effect of the perpendicular anisotropy on the ground-state structure of a 2D spin nanodot in the presence of a dipolar interaction. A core, or region of nonzero magnetization perpendicular to the plane of the system, has been demonstrated to form in the considered nanodot only above a certain threshold value of anisotropy $A$, and this threshold value to grow with the  ratio between $D$, the dipolar coupling, and $J$ the exchange integral.  It also depends on the linear size $L$ of the sample due to the long-range dipolar interaction. In addition, we have observed that with increasing $A$ the regions of nonzero perpendicular magnetization first emerge at the corners of the sample and only afterwards in the center. For very large values of $D$ the GS takes the form of a stripe domain structure; we have analyzed thoroughly the impact of the parameters $A$, $J$ and $D$ on this domain structure. The stripe pattern, namely the number of opposite domains, is shown to depend only on the $D/J$ ratio and to be independent of $A$ in the "perpendicular" regime, namely regime permitted by large enough $A$. This allowed the assignment of each stripe structure to the corresponding range of $D/J$ as shown in Fig. \ref{fig:Stripes}.
From the determined GS's, we have studied  finite-temperature behaviors of nanodots. We found that for small $A/D$, there exists a second-order transition from the in-plane
vortex ordering to the disordered phase: the peak of the specific heat as well as that of the susceptibility strongly depend on the system size. For larger $A/D$,
there is a first-order re-orientation  transition from in-plane to perpendicular ordering with a large latent heat.
To our knowledge, such a transition in a very small system has never been observed before at a finite temperature.
For much larger $A/D$, there is no phase transition.
These results show that the core vortex structure is stabilized at finite temperatures,
making possible applications using perpendicular
magnetization reversal \cite{Shinjo289,Kikuchi90, Xiao102}.

\section*{Acknowledgements}
This work was supported by the Polish National Science Centre (NCN) through the project
UMO-2013/08/M/ST3/00967.

\end{document}